\documentclass[runningheads]{llncs}
\usepackage[pdftex]{graphicx}
\usepackage{epsfig}

\usepackage{tikz}
\usepackage{comment}
\usepackage{amsmath,amssymb} 
\usepackage{cancel}
\usepackage{color}

\usepackage[accsupp]{axessibility}  

\begin{document}
\pagestyle{headings}
\mainmatter
\def\ECCVSubNumber{7188}  

\title{Visual Looming from Motion Field and Surface Normals} 

\titlerunning{Obtaining Visual Looming}
%
\author{Juan Yepes \and
Daniel Raviv}
\authorrunning{J. Yepes and D. Raviv}
%
\institute{Florida Atlantic University, Boca Raton FL 33431, USA \\
\email{\{jyepes, ravivd\}@fau.edu}\\
\url{https://www.fau.edu/engineering/eecs}}
\maketitle

\begin{abstract}
Looming, traditionally defined as the relative expansion of objects in the observer's retina, is a fundamental visual cue for perception of threat and can be used to accomplish collision free navigation. In this paper we derive novel solutions for obtaining visual looming quantitatively from the 2D motion field resulting from a six-degree-of-freedom motion of an observer relative to a local surface in 3D. We also show the relationship between visual looming and surface normals. We present novel methods to estimate visual looming from spatial derivatives of optical flow without the need for knowing range. Simulation results show that estimations of looming are very close to ground truth looming under some assumptions of surface orientations. In addition, we present results of visual looming using real data from the KITTI dataset. Advantages and limitations of the methods are discussed as well.

\keywords{Visual Looming, Optical Flow, Surface Normals.}
\end{abstract}

\section{Introduction}
The visual looming cue, defined quantitatively by \cite{raviv1992quantitative} as the negative instantaneous change of range over the range, is related to the increase in size of an object projected on the observer's retina. Visual looming can be used to define threat regions for obstacle avoidance without the need of range and image understanding. Visual looming is independent of camera rotation and can indicate threat of moving objects as well.

Studies in biology have shown strong evidence of neural circuits in the brains of creatures related to the identification of looming \cite{ache2019neural}. Basically, creatures have instinctive escaping behaviors that tie perception directly to action. This way creatures can avoid impending threat from predators that project an expanding image on their visual systems \cite{evans2018synaptic}\cite{yilmaz2013rapid}. Also, there is evidence of looming being involved in specialized behaviors, for example controlling the action prior to an imminent collision with water surface as exhibited by plummeting gannets \cite{lee1981plummeting} or acrobatic evasive maneuvers exhibited by flies \cite{fabian2022avoiding}\cite{muijres2014flies}.

According to Gibson \cite{gibson2014ecological}, motion relative to a surface is one of the most fundamental visual perceptions. He argued that visual information is processed in a bottom up way, starting from simple to more complex processing. The perceived optical flow resulting from motion of the observer is sufficient to make sense of the environment where a direct connection from visual perception to action can be established. Optical flow analysis is a primitive, simple and robust method for various visual tasks such as distance estimation, image segmentation, surface orientation and object shape estimation \cite{albus1990motion}\cite{cipolla1992surface}.

Optical flow is the estimation of the motion field, which is the 2D perspective projection on the image of the true 3D velocity field as a consequence of the observer's relative motion. Optical flow can be calculated using a number of algorithms that process variations of patterns on image brightness \cite{horn1981determining}\cite{verri1986motion}\cite{zhai2021optical}. 

Optical flow measurements, as the estimation of motion field, includes the necessary information about the relative rate of expansion of objects between two consecutive images from which the visual looming cue can be estimated \cite{raviv1992quantitative}.

In this paper, we present two novel analytical closed form expressions for calculating looming for any six-degree-of-freedom motion, using spatial derivatives of the motion field. The approach can be applied to any relative motion of the camera and any visible surface point. In addition, we show the theoretical relationship between the value of looming and surface orientation. In simulation results we show the effect of the surface normal to the accuracy of the estimated looming values. We demonstrate a new way to extract looming from optical flow using the RAFT model and derived relevant expressions. We show that egomotion is \textit{not }required for estimating looming. In other words, knowledge of range to the point, as well as relative translation or rotation information are \textit{not} required for computing looming and hence for autonomous navigation tasks.

\section{Related Work}

\subsection{Visual Looming}
The visual looming cue is related to the relative change in size of an object projected on the observer's retina as the range to the object increases (Figure \ref{Fig:LoomingCue}.a). It is defined \textit{quantitatively} as the negative value of the time derivative of the range between the observer and a point in 3D divided by the range \cite{raviv1992quantitative}:

\begin{figure}
	\centering
	{\epsfig{file = 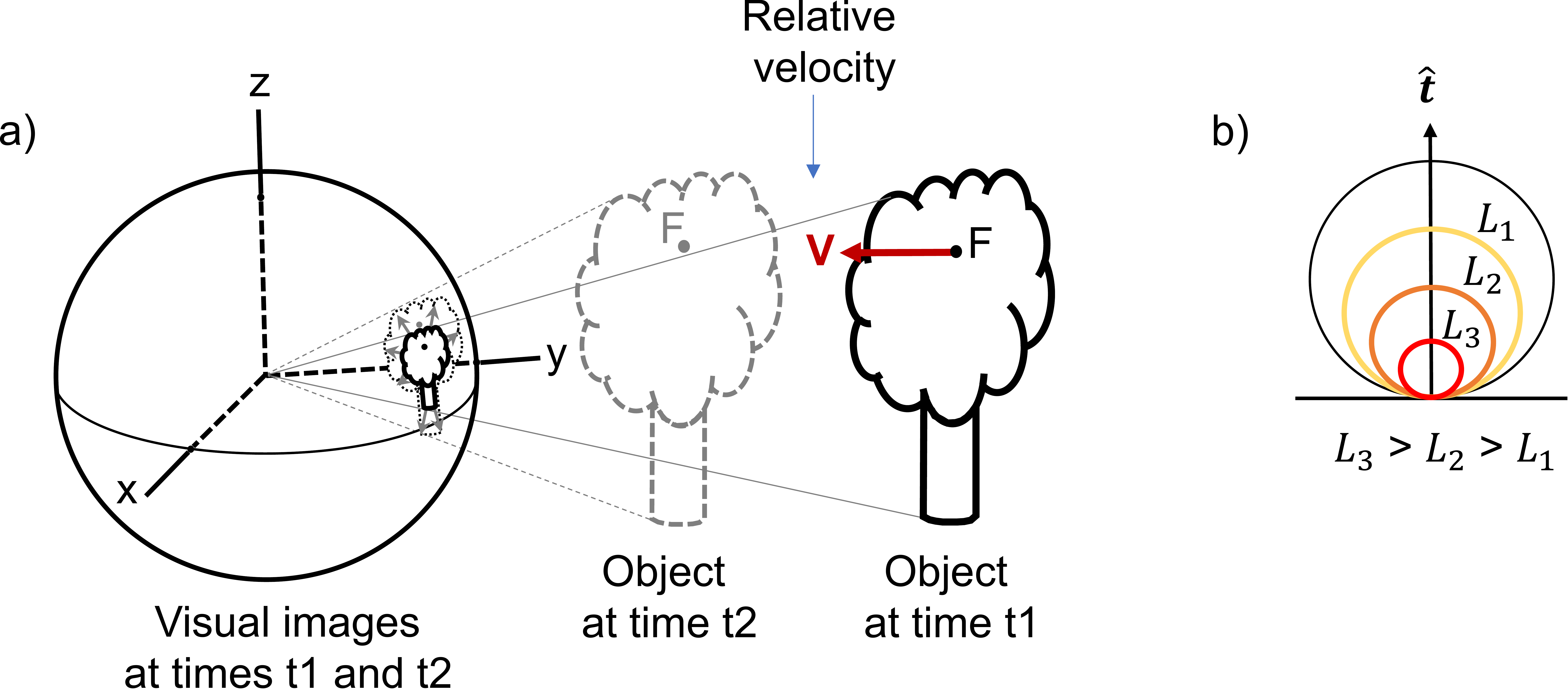, width = 12cm}}
	\caption{Visualizing looming cue: a) relative expansion of 3D object on spherical image, b) equal looming spheres as shown for a given translation $\mathbf{\hat{t}}$}
	\label{Fig:LoomingCue}
\end{figure}

\begin{align}
	L &= - \lim\limits_{\Delta t \to 0}\frac{\left(\frac{r_2-r_1}{\Delta t}\right)}{r_1}  \label{E:LoomingDiscrete}\\	
	L &= -\left(\frac{\dot{r}}{r}\right)  \label{E:Looming}
\end{align}
where $L$ denotes looming, $t_1$ represents time instance 1, $t_2$ represents time instance 2, $\Delta t$ is $t_2 - t_1$, $r_1$ is the range to the point at time instance $t_1$, and $r_2$ is the range at time instance $t_2$. Dot denotes derivative with respect to time.\\

Please note that in this paper we use the terms camera, observer and vehicle interchangeably.

\subsubsection{Looming Properties}
Note that the result for $L$ in equation \eqref{E:Looming} is a scalar value. $L$ is \textit{dependent} on the vehicle translation component but \textit{independent} of the vehicle rotation.
Also $L$ is measured in $[time^{-1}]$ units.

It was shown that points in space that share the same looming values form equal looming spheres with centers that lie on the instantaneous translation vector $\mathbf{t}$ and intersect with the vehicle origin. These looming spheres expand and contract depending on the magnitude of the translation vector \cite{raviv1992quantitative}.

Since an equal looming sphere corresponds to a particular looming value, there are other spheres with varying values of looming with different radii. A smaller sphere signifies a higher value of looming as shown in Figure \ref{Fig:LoomingCue}.b ($L_3 > L_2 > L_1$).

Regions for obstacle avoidance and other behavior-related tasks can be defined using equal looming spheres. For example, a high danger zone for $L>L_3$, medium threat for $L_3 > L > L_2$ and low threat for $L_2 > L > L_1$.

\begin{figure}
	\centering
	{\epsfig{file = 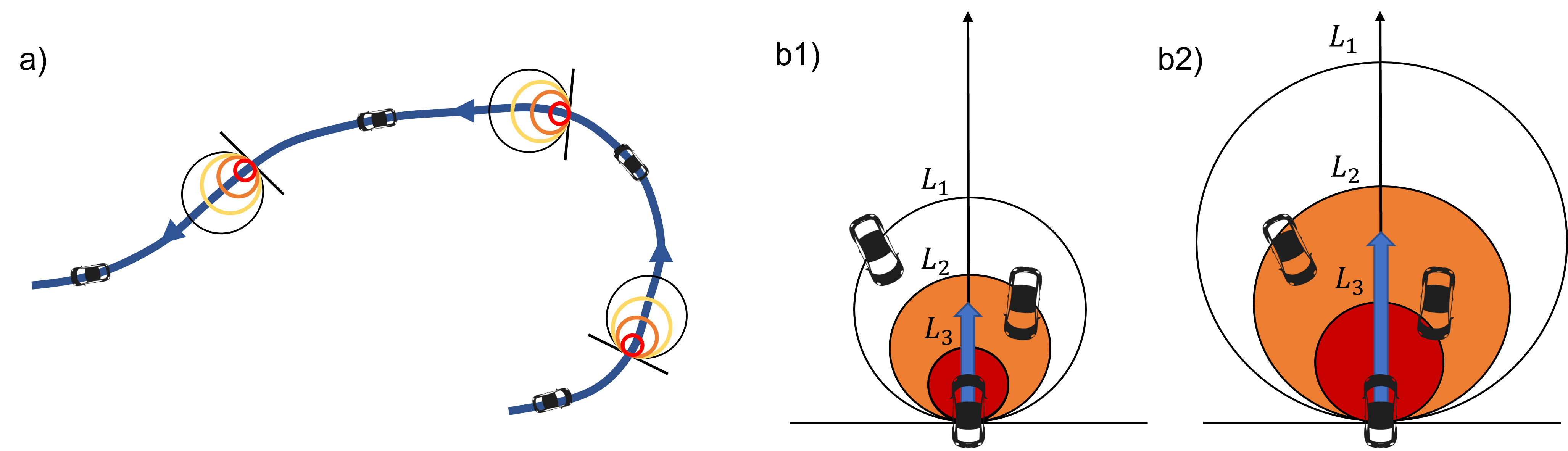, width = 12cm}}
	\caption{Looming spheres: a) Looming spheres superimposed on vehicle trajectory; b) for given identical looming values b1 shows the looming spheres for low speed and b2 shows the looming spheres for high speed}  
	\label{fig:Treat Zones}
\end{figure}

\subsubsection{Advantages of Looming}
 Looming provides time-based imminent threat to the observer caused by stationary environment or moving objects \cite{raviv1992quantitative}. There is no need for scene understanding such as identifying cars, bikes or pedestrians.
 \begin{figure}
 	\centering
 	{\epsfig{file = 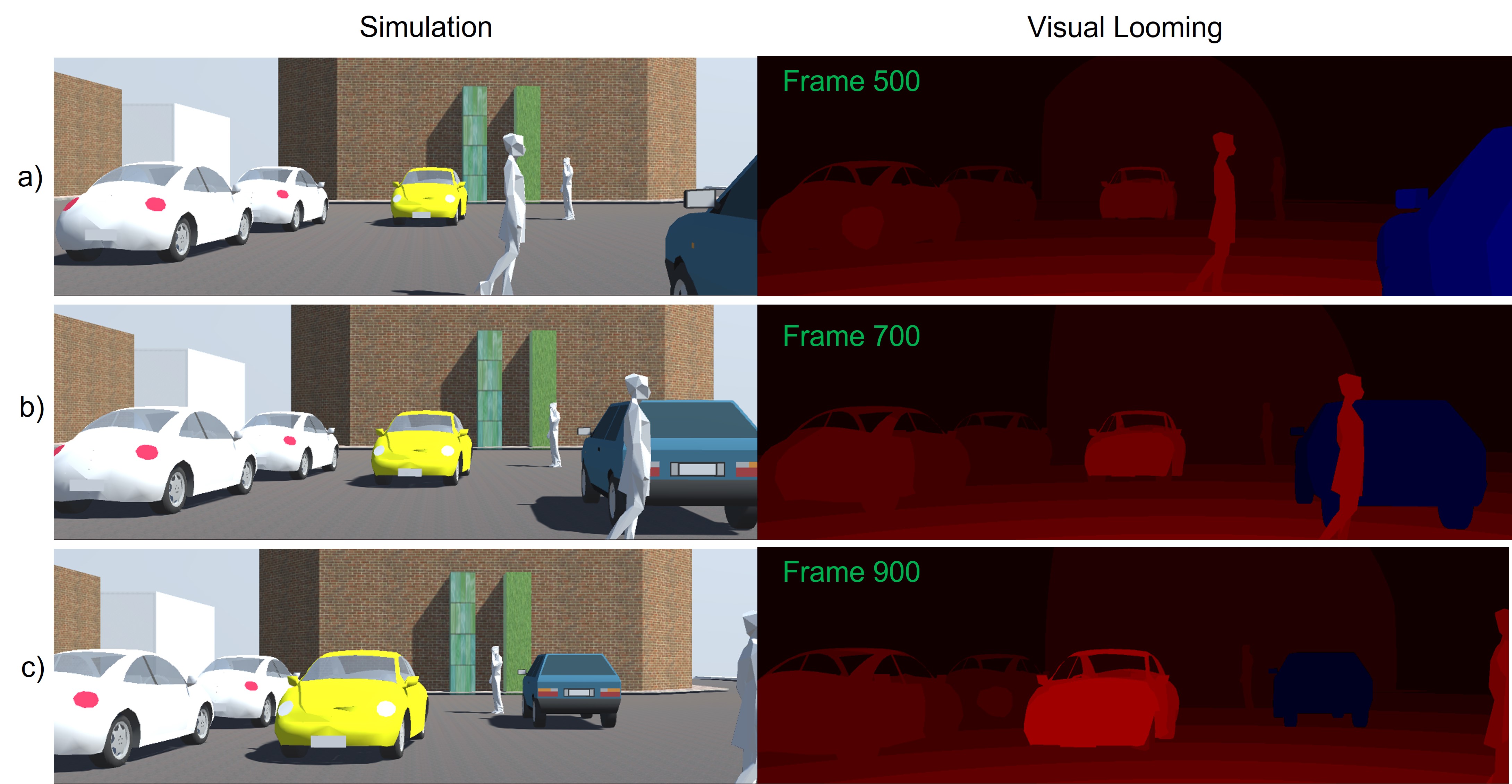, width = 12cm}}
 	\caption{Visual Looming from Simulation: left - a image sequence of the simulated scene. Note that the white and yellow cars are approaching the observer and the blue car is moving away from the observer; right - the corresponding looming values for the image sequence (the brighter the red the higher the threat; blue corresponds to negative looming)}
 	\label{SimulationLooming}
 \end{figure}
Threat regions can be obtained from looming values by assigning specific thresholds. Figure \ref{fig:Treat Zones}.a shows the looming sphere as a function of time for a given constant speed. For the same identical values of looming $L_1$, $L_2$ and $L_3$, Figure \ref{fig:Treat Zones}.b shows equal looming spheres using two (time based) threat zones. Figure \ref{fig:Treat Zones}.b1 shows looming spheres at low speed and Figure \ref{fig:Treat Zones}.b2 shows looming spheres at high speed. Note that the radii of the spheres are proportional to the speed of the vehicle. 

Visual looming can provide indication of threat from moving objects as shown in Figure \ref{SimulationLooming}. Threats from approaching objects are visualized as bright red colors while receding objects appear in blue.

\subsubsection{Measuring Visual Looming}
Several methods were shown to quantitatively extract the visual looming cue from a 2D image sequence by measuring attributes like area, brightness, texture density and image blur \cite{Raviv2000TheVL}.

Similar to visual looming, the Visual Threat Cue (VTC) is a measurable time-based scalar value that provides some measure for a relative change in range between a 3D surface and a moving observer \cite{kundur1999novel}.
Event-based cameras were shown to detect looming objects in real-time from optical flow \cite{ridwan2018looming}.

\subsection{Motion Field and Optical Flow}
\textit{Motion field} is the 2D projection of the true 3D \textit{velocity field} onto the image surface while the \textit{optical flow} is the local apparent motion of points in the image \cite{verri1986motion}. Basically, optical flow is an estimation of the motion field that can be recovered using a number of algorithms that exploit the spatial and temporal variations of patterns of image brightness \cite{horn1981determining}. From the computer vision perspective the question is how to obtain information about the camera motion and objects in the environment from estimations of the motion field \cite{aloimonos1992visual}.
Optical flow algorithms are divided in three categories: knowledge driven, data driven and hybrid. A comprehensive survey of optical flow and scene flow estimation algorithms were provided by \cite{zhai2021optical}. 

In this paper we make use of a state of the art approach to compute optical flow called RAFT (Recurrent all-pairs field transforms for optical flow) \cite{teed2020raft} which we use as \textit{a given input} to our method to obtain estimations of visual looming. Estimation of optical flow from a sequence of images is beyond the scope of this paper.

\section{Method}
In this section we derive two different expressions for visual looming ($L$) for a general six-degrees-of-freedom motion. As mentioned earlier, the camera is attached to the vehicle and both share the same coordinate system. 

\subsection{Velocity and Motion Fields}
\begin{figure}
	\centering
	{\epsfig{file = 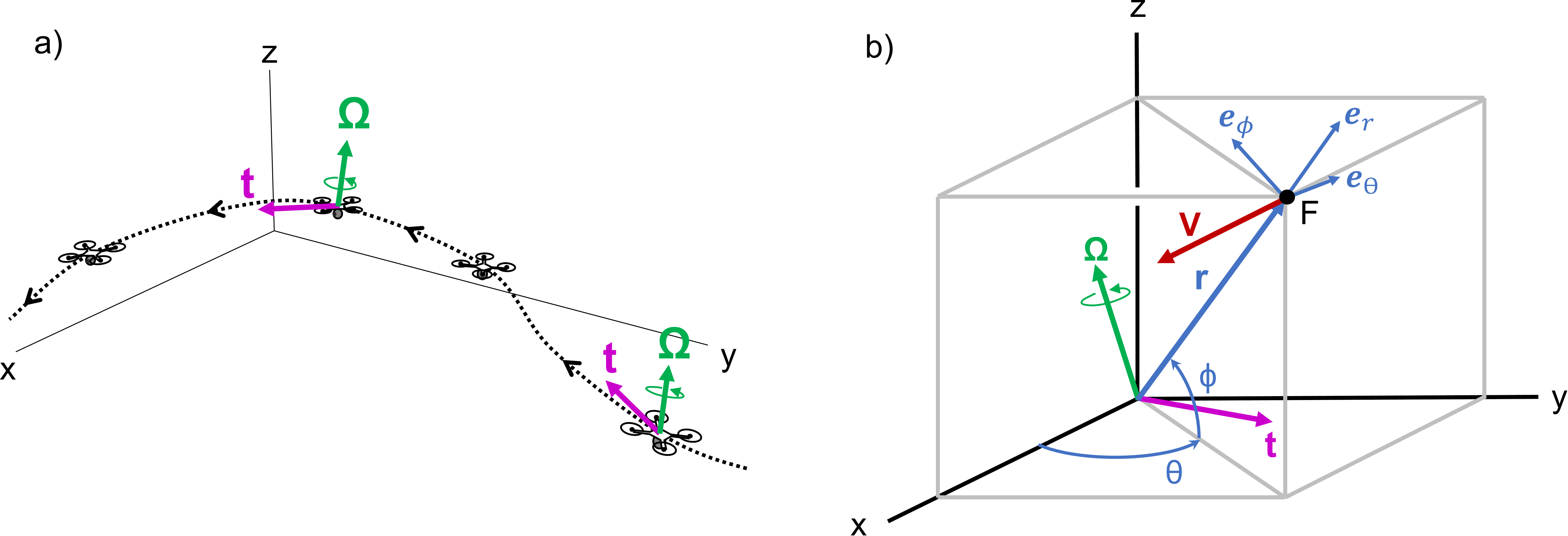, width = 12cm}}
	\caption{General motion of a vehicle in 3D space. a) Vehicle trajectory in world coordinates with translation vector $\mathbf{t}$ and rotation vector $\mathbf{\Omega}$. b) Camera coordinate system}
	\label{fig:VehicleTrajectory}
\end{figure}
Consider a vehicle motion in 3D space relative to an arbitrary stationary reference point $\mathbf{F}$ (refer to Figure \ref{fig:VehicleTrajectory}). At any given time, the vehicle velocity vector consists of a translation component $\mathbf{t}$ and a rotation component $\mathbf{\Omega}$ in world coordinates as shown in Figure \ref{fig:VehicleTrajectory}.a. Also consider a local coordinate system centered at the camera which is fixed to the moving vehicle (Figure \ref{fig:VehicleTrajectory}.b). We choose the z-axis to be aligned with the vertical orientation of the camera and the x-axis to be aligned with the optical axis of the camera. In this frame any stationary point in the 3D scene can be represented using spherical coordinates $(r,\theta,\phi)$ where $r$ is the range to the point $\mathbf{F}$, $\theta$ is the azimuth angle measured from the x-axis and $\phi$ is the elevation angle from the XY-plane.
 
In camera coordinates, a point $\mathbf{F}$ has an associated relative velocity $\mathbf{V}$ with a translation vector $\mathbf{-t}$ and rotation vector $\mathbf{-\Omega}$, given by (see \cite{meriam2012engineering}): 
\begin{align}
	\mathbf{V} &= (-\mathbf{t}) + (-\mathbf{\Omega} \times \mathbf{r}) \label{E:VelocityField} 
\end{align}
Note that $\mathbf{r}$ in bold refers to the range vector and the scalar $r$ refers to its magnitude, i.e., $r = |\mathbf{r}|$. 
Notice also that for a given point $\mathbf{F}$, the velocity vector $\mathbf{V}$ is the \textit{velocity field} in 3D due to egomotion of the camera.

\begin{flushleft}
	By dividing equation \eqref{E:VelocityField} by the scalar $r$ and expanding $\mathbf{t}$ and $\mathbf{\Omega}$ we get:
\end{flushleft}
\begin{align}
	\frac{\mathbf{V}}{r} &= 
	\left(\frac{-\mathbf{t}}{r}\right) + \left(\frac{\mathbf{-\mathbf{\Omega} \times \mathbf{r}}}{r}\right)\notag\\
	&=\frac{-(t_r\mathbf{e}_r + t_\theta\mathbf{e}_\theta + t_\phi\mathbf{e}_\phi )}{r} - (\Omega_r\mathbf{e}_r + \Omega_\theta\mathbf{e}_\theta + \Omega_\phi\mathbf{e}_\phi ) \times \mathbf{e}_r \notag\\
	&=\left(\frac{-t_r}{r}\right)\mathbf{e}_r + \left(\frac{-t_\theta}{r}\right)\mathbf{e}_\theta + 
	\left(\frac{-t_\phi}{r}\right)\mathbf{e}_\phi + 
	(-\Omega_\phi) \mathbf{e}_\theta +
	\Omega_\theta \mathbf{e}_\phi \label{E:VelocityFieldLeft}
\end{align}
where:
\begin{align}
	t_r &= \mathbf{t}\cdot \mathbf{e}_r ,&&\Omega_r = \mathbf{\Omega}\cdot \mathbf{e}_r\notag\\
	t_\theta &= \mathbf{t}\cdot \mathbf{e}_\theta ,&&\Omega_\theta = \mathbf{\Omega}\cdot \mathbf{e}_\theta\notag\\
	t_\phi &= \mathbf{t}\cdot \mathbf{e}_\phi ,&&\Omega_\phi = \mathbf{\Omega}\cdot \mathbf{e}_\phi\notag
\end{align}

The \textit{velocity field} $\mathbf{V}$ can also be expressed in spherical coordinates $(r,\theta,\phi)$ and directional unit vectors ($\mathbf{e}_r$,$\mathbf{e}_\theta$,$\mathbf{e}_\phi$) as shown in \cite{meriam2012engineering}:
\begin{align}
	\mathbf{V} &= \dot{r}\mathbf{e}_r + r\dot{\theta}\cos(\phi)\mathbf{e}_\theta + r\dot{\phi}\mathbf{e}_\phi \label{E:VelocityFieldSpherical}
\end{align}
By dividing equation \eqref{E:VelocityFieldSpherical} by $r$ we obtain:
\begin{equation}
	\frac{\mathbf{V}}{r}
	= \left(\frac{\dot{r}}{r}\right)\mathbf{e}_r + 
	\dot{\theta}\cos(\phi)\mathbf{e}_\theta + 
	\dot{\phi}\mathbf{e}_\phi\label{E:VelocityFieldSphericalRight}
\end{equation}
From \eqref{E:VelocityFieldSphericalRight} we can identify the 2D \textit{motion field} $\mathbf{f}$ defined as:
\begin{equation}
	\mathbf{f}
	= \dot{\theta}\cos(\phi)\mathbf{e}_\theta + 
	\dot{\phi}\mathbf{e}_\phi\label{E:MotionField}
\end{equation}

Note that $\left(\frac{\dot{r}}{r}\right)$ is not part of the motion field since it is along the range direction $\mathbf{e}_r$  from the camera to point $\mathbf{F}$. The \textit{motion field} $\mathbf{f}$ in expression  \eqref{E:MotionField} is the projection of the \textit{velocity field} $\mathbf{V}$ on the spherical image.

\subsection{Looming from Spatial Partial Derivatives of the Velocity Field}

The Looming value $(L)$ is related to the relative expansion of objects projected on the image of the camera. This means that there is a direct relationship between the change in the \textit{motion field} in the vicinity of a point and looming $(L)$. In order to find this relationship we apply spatial partial derivatives with respect to $\theta$ and $\phi$ to the velocity field divided by r as described in equations \eqref{E:VelocityFieldLeft} and \eqref{E:VelocityFieldSphericalRight} to get:
\begin{align}
	\frac{\partial}{\partial \theta}\left(\frac{\mathbf{V}}{r}\right) &=\frac{\partial}{\partial \theta}\left[\left(\frac{-t_r}{r}\right)\mathbf{e}_r + \left(\frac{-t_\theta}{r}\right)\mathbf{e}_\theta + 
	\left(\frac{-t_\phi}{r}\right)\mathbf{e}_\phi + 
	(-\Omega_\phi) \mathbf{e}_\theta +
	\Omega_\theta \mathbf{e}_\phi\right]\notag\\
	&=\frac{\partial}{\partial \theta}\left[\left(\frac{\dot{r}}{r}\right)\mathbf{e}_r + 
	\dot{\theta}\cos(\phi)\mathbf{e}_\theta + 
	\dot{\phi}\mathbf{e}_\phi\right] \label{E:VPartialDerivativesTheta}
\end{align}
\begin{align}
	\frac{\partial}{\partial \phi}\left(\frac{\mathbf{V}}{r}\right) &=\frac{\partial}{\partial \phi}\left[\left(\frac{-t_r}{r}\right)\mathbf{e}_r + \left(\frac{-t_\theta}{r}\right)\mathbf{e}_\theta + 
	\left(\frac{-t_\phi}{r}\right)\mathbf{e}_\phi + 
	(-\Omega_\phi) \mathbf{e}_\theta +\Omega_\theta \mathbf{e}_\phi \right] \notag\\
	&=\frac{\partial}{\partial \phi}\left[\left(\frac{\dot{r}}{r}\right)\mathbf{e}_r + 
	\dot{\theta}\cos(\phi)\mathbf{e}_\theta + 
	\dot{\phi}\mathbf{e}_\phi\right]
	\label{E:VPartialDerivativePhi}
\end{align}
\begin{flushleft}
By performing the corresponding derivations\footnote{Most of the detailed derivations were omitted from the paper due to page limit. The detailed derivations are available upon request.} for equations \eqref{E:VPartialDerivativesTheta} and \eqref{E:VPartialDerivativePhi}, in addition to using equation \eqref{E:Looming} for $L$, we obtain the following two independent expressions for looming ($L$):
\end{flushleft}
\begin{align}
	L & = \frac{\partial \dot{\theta}}{\partial \theta} - \dot{\phi}\tan(\phi) - \frac{t_\theta}{r}\left(\frac{1}{\cos(\phi)}\right)\left(\frac{1}{r}\frac{\partial r}{\partial \theta}\right)\label{E:LoomingFromTheta}	
\end{align}	
\begin{align}
	L &= \frac{\partial\dot{\phi}}{\partial \phi} -\frac{t_\phi}{r}\left(\frac{1}{r}\frac{\partial r}{\partial\phi}\right)\label{E:LoomingFromPhi} 
\end{align}

In both expressions, $L$ is a scalar value that can be computed from the spatial change in the \textit{motion field}. Note that $L$ is independent of the vehicle  rotation $\mathbf{\Omega}$ and is dependent only on the translation components scaled by $r$, $t_\theta/r$ and $t_\phi/r$.  

These expressions may also apply to any relative motion of a point in 3D for any six-degrees-of-freedom motion.

\subsection{Estimation of Visual Looming}
Expressions \eqref{E:LoomingFromTheta} and \eqref{E:LoomingFromPhi} contain $r$, which is not measurable from the image. If we estimate looming by eliminating the components that contain $r$ in \eqref{E:LoomingFromTheta} and \eqref{E:LoomingFromPhi} then estimates for $L$ are obtained as:

\begin{equation}\label{E:LoomingEstimatedThetaDot}
	L_{est1} = \frac{\partial \dot{\theta} }{\partial \theta} - \dot{\phi}\tan{\phi}
\end{equation}

\begin{equation}\label{E:LoomingEstimatedPhiDot}
	L_{est2} = \frac{\partial \dot{\phi} }{\partial \phi} 
\end{equation}
Later in the paper, we show the effect of eliminating the components in equations \eqref{E:LoomingEstimatedThetaDot} and \eqref{E:LoomingEstimatedPhiDot} that include $r$ and its derivative on the error in calculating looming.

Notice that $L_{est1}$ and $L_{est2}$ can be obtained directly from measurements of the horizontal and vertical components of the motion field for a particular time instance, specifically, the changes of the values of the motion field in the spatial dimensions $\theta$ and $\phi$. This is the meaning of the spatial partial derivatives $\frac{\partial \dot{\theta} }{\partial \theta}$, $\frac{\partial \dot{\phi} }{\partial \phi}$ in expressions \eqref{E:LoomingEstimatedThetaDot} and \eqref{E:LoomingEstimatedPhiDot}.

\subsection{Surface Normals}
\label{S:SurfaceNormals}
\begin{figure}
	\centering
	{\epsfig{file = 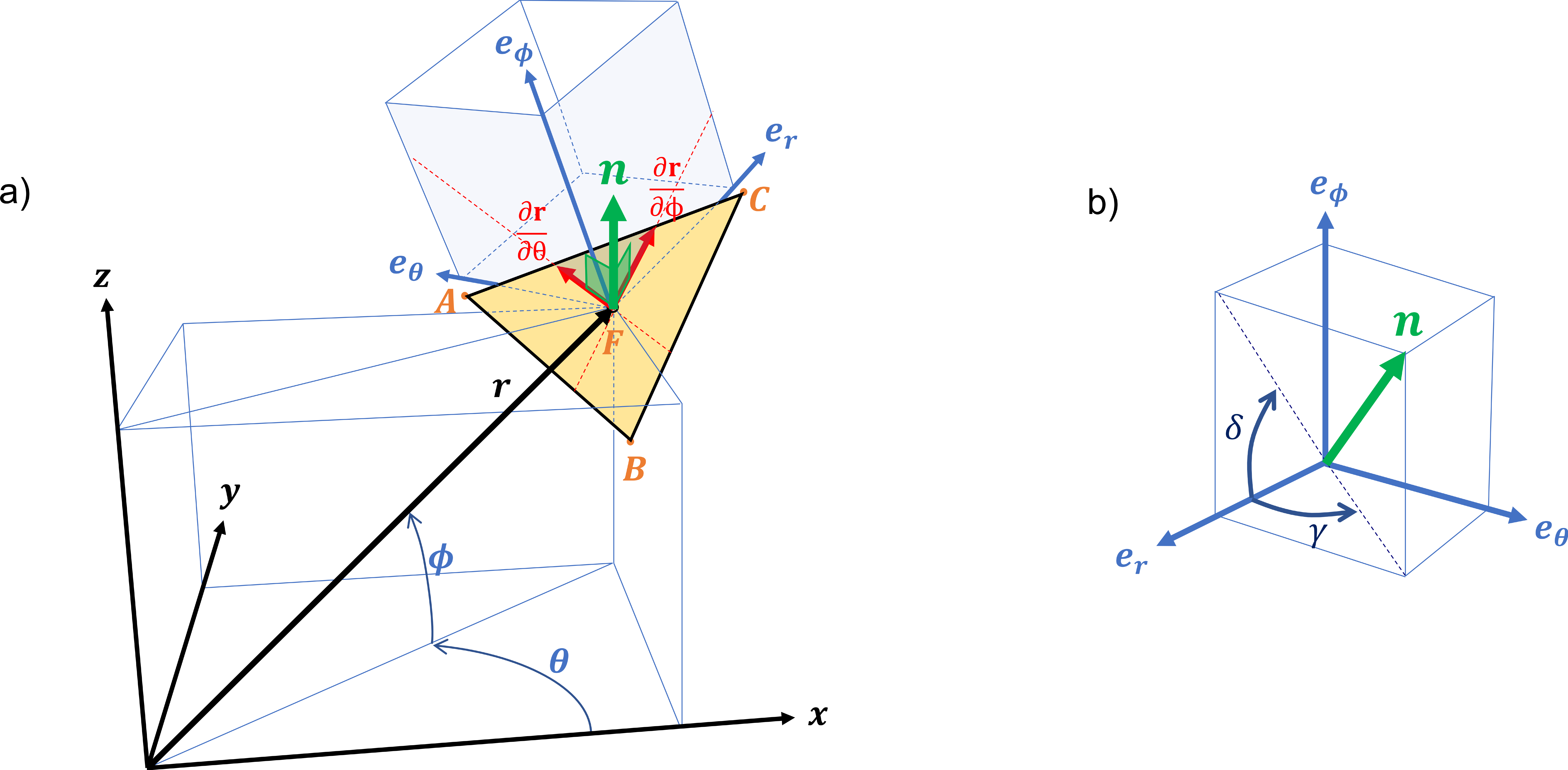, width = 12cm}}
	\caption{Surface orientation: a) infinitesimal planar surface represented by the triangle ABC; b) tilt angles related to the normal of the planar patch}
	\label{fig:TiltAngles}
\end{figure}

The point $\mathbf{F}$ lies on a infinitesimally small surface patch with normal vector $\mathbf{n}$. To find the relation between $L$ and $\mathbf{n}$ we assume that the surface is planar in the vicinity of the point. This planar patch is represented by the triangle ABC and $\mathbf{n}$ is the normal at point $\mathbf{F}$ (see Figure \ref{fig:TiltAngles}.a). 

We can identify two vectors, $\frac{\partial \mathbf{r}}{\partial\theta}$ and $\frac{\partial \mathbf{r}}{\partial\phi}$, located on the planar patch, that result from the displacement of $\mathbf{r}$ along the angles $\theta$ and $\phi$. Notice that these vectors are both perpendicular to the surface normal $\mathbf{n}$ and are located at the intersection of the planes ABC,   $\mathbf{e}_r\mathbf{e}_\phi$, and $\mathbf{e}_\theta\mathbf{e}_\phi$. It can be shown that the following constraints hold for any infinitesimal small patch:
\begin{align}
	\frac{\partial \mathbf{r}}{\partial\theta}\cdot\mathbf{n} &= 0\label{E:rDotconstrainTheta}\\
	\frac{\partial \mathbf{r}}{\partial\phi}\cdot\mathbf{n} &= 0\label{E:rDotconstrainPhi}
\end{align}
Using $\mathbf{r} = r\mathbf{e}_r$ and solving for $\frac{\partial\mathbf{r}}{\partial\theta}$ and $\frac{\partial\mathbf{r}}{\partial\phi}$ in \eqref{E:rDotconstrainTheta} and \eqref{E:rDotconstrainPhi} we obtain:

\begin{align}
	\frac{1}{r}\frac{\partial r}{\partial\theta} &=  -\cos\phi\left(\frac{\mathbf{e}_\theta\cdot \mathbf{n}}{\mathbf{e}_r\cdot\mathbf{n}}\right)\label{E:rDotOverrThetaNormal}
\end{align}

\begin{align}
	\frac{1}{r}\frac{\partial r}{\partial\phi} &=  -\left(\frac{\mathbf{e}_\phi\cdot \mathbf{n}}{\mathbf{e}_r\cdot\mathbf{n}}\right)\label{E:rDotOverrPhiNormal}
\end{align}
By defining the surface tilt angles as $\gamma$ and $\delta$ (see Figure \ref{fig:TiltAngles}.b) we obtain:
\begin{align}
	\gamma &= \tan^{-1}\left(\frac{\mathbf{e}_\theta\cdot \mathbf{n}}{\mathbf{e}_r\cdot\mathbf{n}}\right)\label{E:tanGamma}\\
	\delta &= \tan^{-1}\left(\frac{\mathbf{e}_\phi\cdot \mathbf{n}}{\mathbf{e}_r\cdot\mathbf{n}}\right) \label{E:tanDelta}
\end{align}
We can rewrite equations \eqref{E:LoomingFromTheta}, \eqref{E:LoomingFromPhi} using \eqref{E:rDotOverrThetaNormal}, \eqref{E:rDotOverrPhiNormal}, \eqref{E:tanGamma} and \eqref{E:tanDelta} as:

\begin{align}
	L &= \frac{\partial \dot{\theta} }{\partial \theta} - \dot{\phi}\tan{\phi} + \left(\frac{t_\theta}{r}\right)\tan\gamma
	\label{E:Loominggamma}\\	
	L &= \frac{\partial \dot{\phi} }{\partial \phi} + \left(\frac{t_\phi}{r}\right)\tan\delta\label{E:Loomingdelta}
\end{align}
Equations \eqref{E:Loominggamma} and \eqref{E:Loomingdelta} are essentially the same as expressions \eqref{E:LoomingFromTheta} and \eqref{E:LoomingFromPhi} but using normal notations.

\section{Looming Results}
We present two sets of quantitative results of the methods for estimation of looming using simulations and real data.

\subsection{Looming from Simulation}

\begin{figure}
	\centering
	{\epsfig{file = 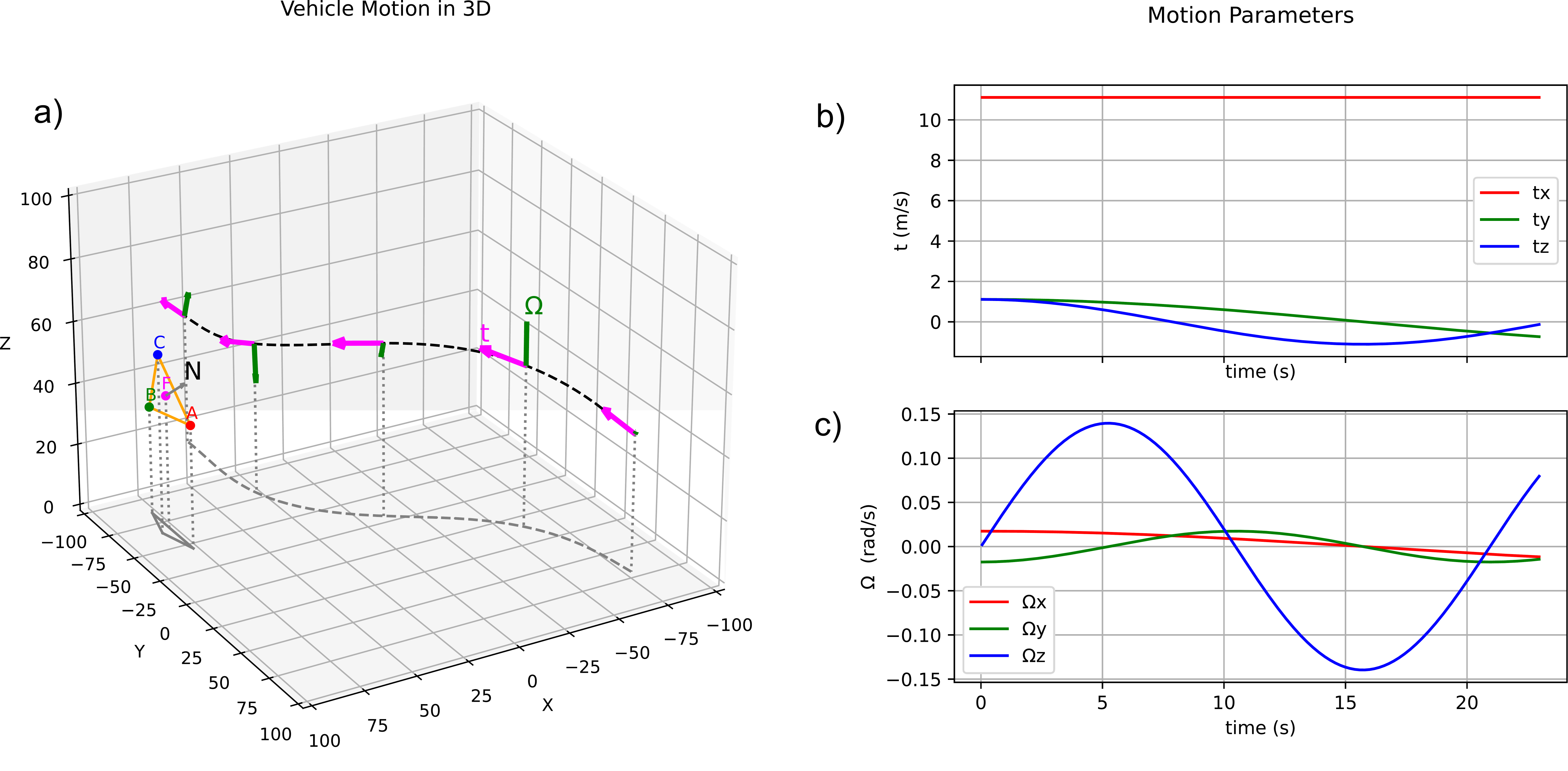, width = 12cm}}
	\caption{Simulation of a vehicle moving in 3D space and a point on a planar patch. a) Vehicle trajectory. b) Motion parameters $\mathbf{t}$ and $\mathbf{\Omega}$ }
	\label{fig:PlanarPatch}
\end{figure}

We simulated a translating and rotating observer in 3D. Measurements were taken for a single stationary point on a tilted planar patch (see Figure \ref{fig:PlanarPatch}.a).

The simulation duration was 23 seconds and samples were taken at 60 Hz (1380 samples in total). The vehicle starting position was $\mathbf{P} = -75\mathbf{i} + 75\mathbf{j}  + 44.3\mathbf{k}$ and the orientation was [forward, left, up] =  $[-\mathbf{i}, -\mathbf{j}, \mathbf{k}]$. The planar patch was simulated by the following points: 
$\mathbf{A} =  80\mathbf{i} - 40\mathbf{j} + 40\mathbf{k}$, 
$\mathbf{B} =  80\mathbf{i} - 80\mathbf{j} + 35\mathbf{k}$ and
$\mathbf{C} =  85\mathbf{i} - 60\mathbf{j} + 58\mathbf{k}$. All distances were in meters.
The vehicle was moving at a speed of s = 11.11 m/s (or 40 km/h) with translation and rotation vectors
$\mathbf{t} = s\mathbf{i} + 0.1s\cos(0.1t)\mathbf{j} + 0.1s\cos(0.2t)\mathbf{k} $ and 
$\mathbf{\Omega} = \cos(0.1t)\mathbf{i} - \cos(0.3t)\mathbf{j} + 8(\frac{\pi}{180})\sin(0.3t)\mathbf{k} $. 

Plots of the parameters $\mathbf{t}$ and $\mathbf{\Omega}$ over time are shown on Figure \ref{fig:PlanarPatch}.b.

\begin{figure}
	\centering
	{\epsfig{file = 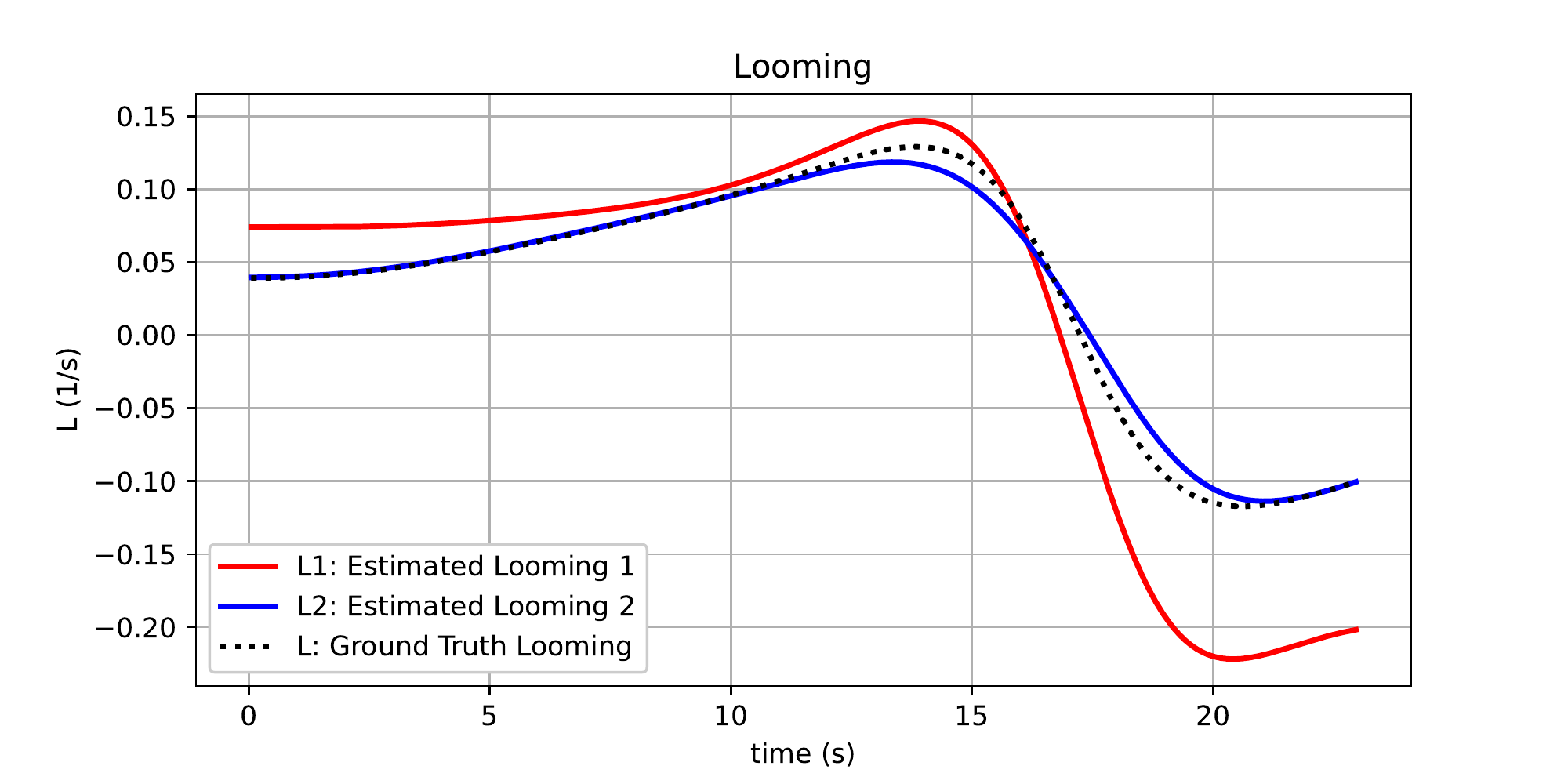, width = 12cm}}
	\caption{Ground truth looming ($L$) and estimated looming ($L_1$ and $L_2$)}
	\label{fig:Looming1}
\end{figure}

\begin{figure}
	\centering
	{\epsfig{file = 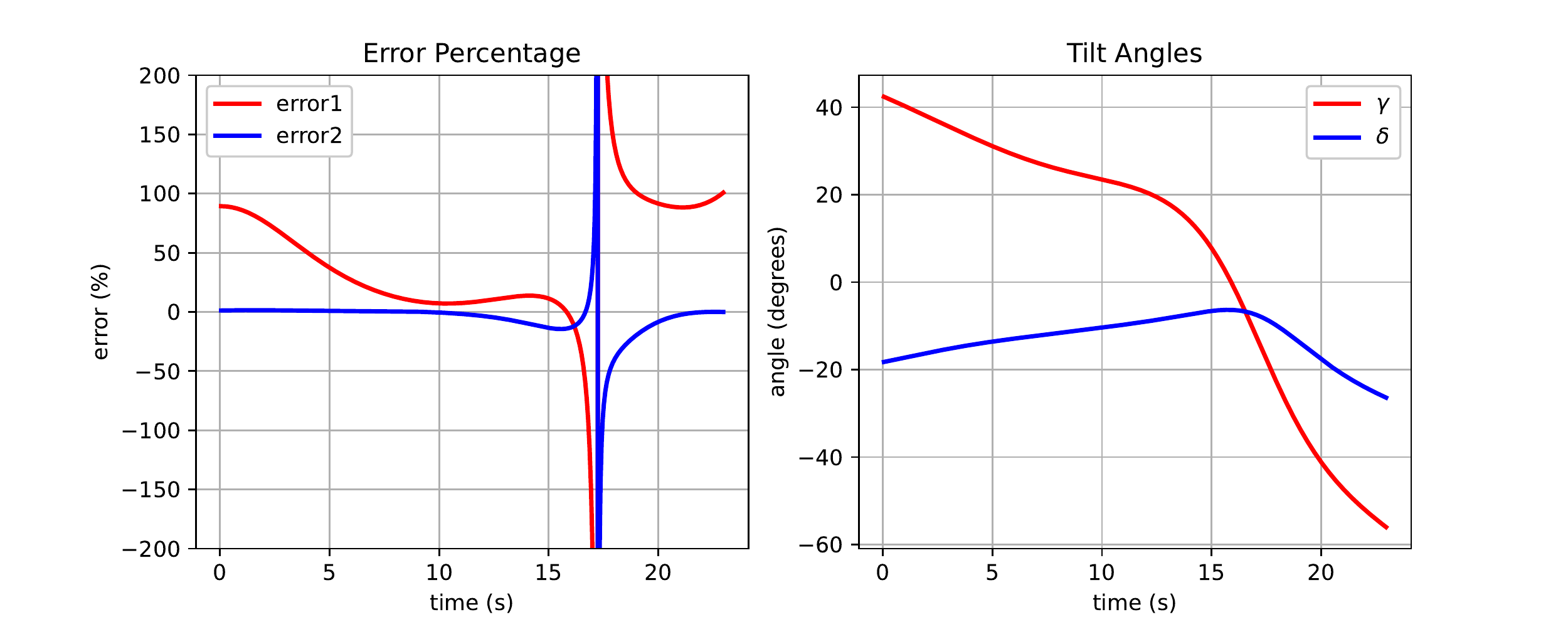, width = 12cm}}
	\caption{Left: error percentage ($error_1$, $error_2$) between ground truth $L$ and estimated looming values $L_1$, $L_2$. Right: tilt angles $\gamma$ and $\delta$ related to the normal of the planar patch}
	\label{fig:error}
\end{figure}
Ground truth looming $L$ is computed from range and its time derivative using equation \eqref{E:LoomingDiscrete}. In addition, two estimations of looming $L_1$ and $L_2$ are computed using equations \eqref{E:Loominggamma} and \eqref{E:Loomingdelta} (see Figure \ref{fig:Looming1}).  
To compare $L$ with each estimate of $L_1$ or $L_2$ we use the error percentage metric: 
\begin{align}
	error_i (\%) &=  \frac{L_i - L}{L}\times 100 &&\text{for } i = 1,2 \label{E:errorMetric}
\end{align}

Plots, as a function of time, for both error values: $error_1$ and $error_2$ that correspond to $L_1$ and $L_2$ are shown in Figure \ref{fig:error} (left). In Figure \ref{fig:error} (right), values of the tilt angles of the planar patch $\gamma$ and $\delta$ are presented. 

Notice that for this particular simulation, $L_2$ values are closer to $L$ than $L_1$ values. This can be explained by a smaller value of the tilt angle $\delta$ which introduces less estimation error. The discontinuity in the error signal plot (around time = 17.2s) is due to $L$ becoming closer to zero in equation \eqref{E:errorMetric}.

For example, in Figure \ref{fig:error} the ground truth looming has a positive maximum $L = 0.129 s^{-1}$ at $t = 13.8s$, and the estimated values are $L_1 = 0.147s^{-1}$, and $L_2 = 0.117s^{-1}$. Both estimates get very close to the ground truth of $L$ with $error_1 = 13\%$ and $error_2 = -9\%$. 

This simulation shows that when tilt angles are small (lower than $\pm$20 degrees) the error in the estimation of looming is within $\pm$15$\%$ range. Under these assumptions looming estimates are good enough to define threat zones for collision avoidance tasks where thresholds can be adjusted by this margin of error. 

\subsection{Looming from Real Data}
\begin{figure}
	\centering
	{\epsfig{file = 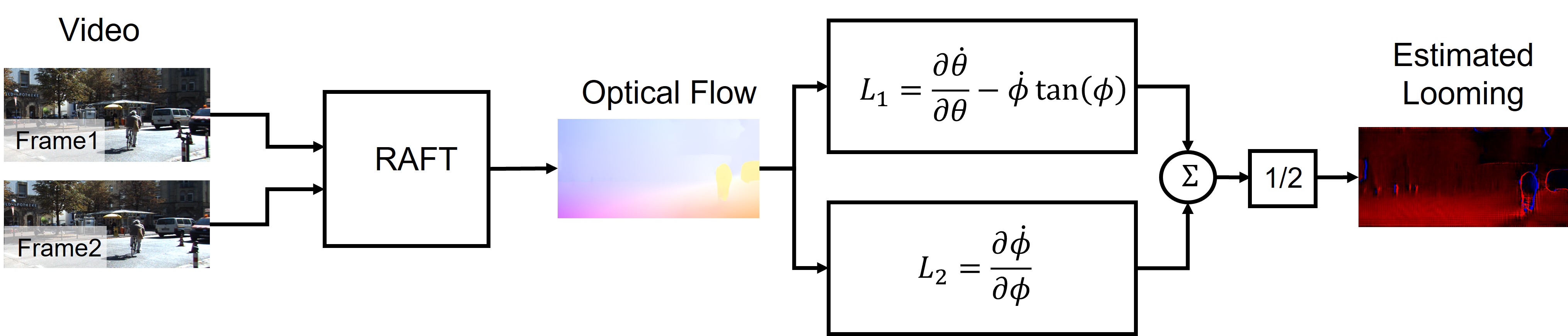, width = 12cm}}
	\caption{Estimation of looming using real data}
	\label{fig:BlockDiagram}
\end{figure}

Visual looming estimates were obtained from optical flow (estimation of the motion field) from a sequence of images taken from a moving camera fixed on a vehicle. The block diagram in Figure \ref{fig:BlockDiagram} shows the process and the different components. 

\begin{figure}
	\centering
	{\epsfig{file = 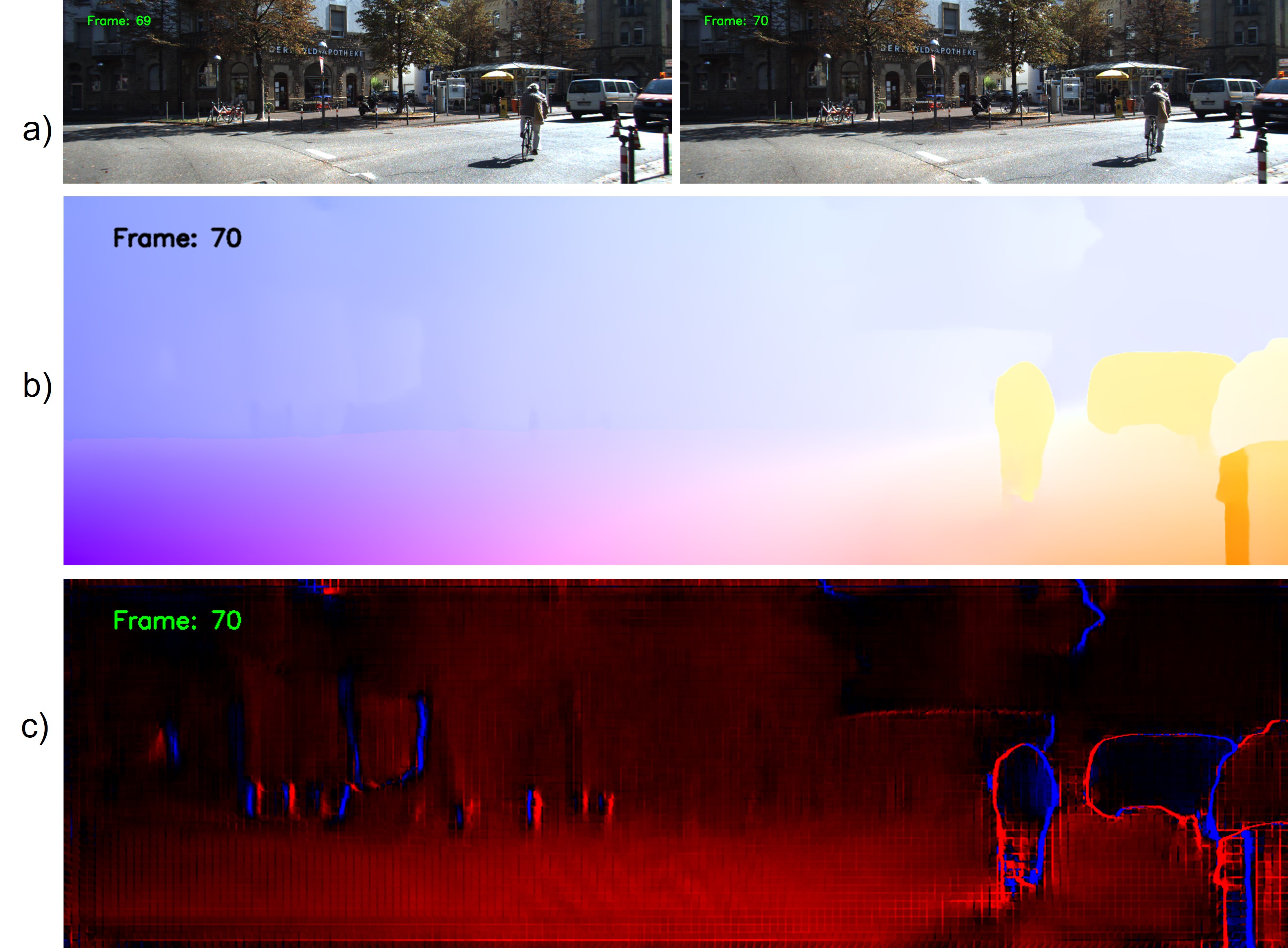, width = 12cm}}
	\caption{Estimated Looming from KITTI. a) Original images (frame 69 and frame 70), b) Optical flow estimate from RAFT. c) Estimated Looming using the average value from equations \eqref{E:LoomingEstimatedThetaDot} and \eqref{E:LoomingEstimatedPhiDot}}
	\label{fig:LoomingFromKitti}
\end{figure}
We processed real data using a particular city drive from the well known KITTI dataset \cite{geiger2013vision}. This dataset includes raw data from several sensors mounted on a vehicle. We used a video sequence from one of the color cameras (FL2-14S3C-C, 1.4 Megapixels, 1392x512 pixels). Then a pair of consecutive RGB images from this dataset were processed by a deep neural network called RAFT (Recurrent All-Pairs Field Transforms) \cite{teed2020raft}. RAFT provides optical flow estimation as output, for each pixel in the image as a displacement pixel pair $(u, v)$. 

We used the KITTI camera intrinsic parameters to transform from pixel image coordinates to spherical coordinates and obtained ($\theta, \phi, \dot{\theta}, \dot{\phi}$). Then the average of spatial derivatives from equations \eqref{E:LoomingEstimatedThetaDot} and \eqref{E:LoomingEstimatedPhiDot} were applied to estimate and visualize the looming values.

Looming estimation from frames 69 and 70 are shown in Figure \ref{fig:LoomingFromKitti}.c. Positive values of looming are shown in red and negative values in blue. The brighter the red, the higher the looming threat. 

Notice an \textit{edge effect} around objects with high values of red and blue colors. This is due to occlusions that generate sudden changes in optical flow and cause distortion of looming, resulting in incorrect values. Those may be managed by discarding some blue regions using additional filtering.

Notice also some noise, shown as square artifacts, due the up-sampling interpolation stage of the RAFT algorithm (1:8 ratio). The noise is due to spatial partial derivatives that have a tendency to amplify minor differences of optical flow.

Moving objects, like the bike and the white car in the image sequence, are shown with darker red color (meaning less threat) due to small relative velocity with respect to the camera. This is an advantage of the approach where estimation of looming is able to catch the relevant threat of moving objects. This is clearly an advantage of looming over conventional perception of depth.

\section{Conclusions}
Visual looming has been shown to be an important cue for navigation and control tasks. This paper explores new ways for deriving and estimating visual looming for general six-degrees-of-freedom motion. The main contributions of the paper are:
\begin{enumerate}
\item We derived two novel analytical closed-form expressions for calculating looming for any six-degree-of-freedom motion. These expressions include spatial derivatives of the motion field. This approach can be applied to any relative motion between an observer and any visible point. 
\item We showed the theoretical relationship of surface normal to the values of looming.
\item We presented simulation results of the effect of surface normals on values of calculated looming. Quantitative calculations showed the relationship between the angles of the surface normal relative to the angle of the range vector and the related effects on the accuracy of estimating looming values.
\item We demonstrated how to extract looming from optical flow. The output of the RAFT model was used to estimate looming values using expressions that were derived in the paper.
\end{enumerate}
It should also be emphasized that knowledge of range to the 3D point, translation or rotation of the camera (egomotion) is \textit{not} required for the estimation of looming and hence for autonomous navigation tasks. 

\section*{Acknowledgments}
The authors thank \textcolor{black}{Dr. Sridhar Kundur} for many fruitful discussion and suggestions as well as very detailed comments and clarifications that led to meaningful improvements of this paper. The authors also thank Nikita Ostro and Benjamin Thaw for their thorough review of this paper.

\clearpage
%
%
\bibliographystyle{splncs04}
\bibliography{LidarLooming.bib}

\title{Supplementary Material: \\Visual Looming from Motion Field and Surface Normals} 
\titlerunning{Obtaining Visual Looming}

\author{Juan Yepes \and
	Daniel Raviv}

\authorrunning{J. Yepes and D. Raviv}

\institute{Florida Atlantic University, Boca Raton FL 33431, USA \\
	\email{\{jyepes, ravivd\}@fau.edu}\\
	\url{https://www.fau.edu/engineering/eecs}}

\maketitle

\setcounter{secnumdepth}{4}
\setcounter{section}{0}
\setcounter{equation}{0}
\setcounter{figure}{0}
\setcounter{table}{0}
\setcounter{page}{1}
\makeatletter
\renewcommand{\thesection}{S\arabic{section}}
\renewcommand{\thesubsubsection}{\alph{subsubsection}{)}}
\renewcommand{\theequation}{S\arabic{equation}}
\renewcommand{\thefigure}{S\arabic{figure}}
\renewcommand{\thetable}{S\arabic{table}}

\section{Introduction}
This supplementary material describes detailed derivations for visual looming $(L)$ using a method that applies spatial partial derivatives $ \frac{\partial ()}{\partial \theta}, \frac{\partial ()}{\partial \phi}$  to the \textit{normalized relative velocity field} $(\frac{\mathbf{V}}{r})$.\\
\section{Derivations}
\subsection{Velocity Field $V$:}
In camera coordinates a point $\mathbf{F}$ has an associated relative velocity $\mathbf{V}$ with a translation vector $\mathbf{-t}$ and rotation vector $\mathbf{-\Omega}$, given by: 
\begin{align}
	\mathbf{V} &= (-\mathbf{t}) + (-\mathbf{\Omega} \times \mathbf{r}) \label{E:SVelocityField} 
\end{align}
By dividing equation \eqref{E:SVelocityField} by the scalar $r$ and expanding $\mathbf{t}$ and $\mathbf{\Omega}$ we get:
\begin{align}
	\frac{\mathbf{V}}{r} &= 
	\left(\frac{-\mathbf{t}}{r}\right) + \left(\frac{\mathbf{-\mathbf{\Omega} \times \mathbf{r}}}{r}\right)\notag\\
	&=\frac{-(t_r\mathbf{e}_r + t_\theta\mathbf{e}_\theta + t_\phi\mathbf{e}_\phi )}{r} - (\Omega_r\mathbf{e}_r + \Omega_\theta\mathbf{e}_\theta + \Omega_\phi\mathbf{e}_\phi ) \times \mathbf{e}_r \notag\\
	&=\left(\frac{-t_r}{r}\right)\mathbf{e}_r + \left(\frac{-t_\theta}{r}\right)\mathbf{e}_\theta + 
	\left(\frac{-t_\phi}{r}\right)\mathbf{e}_\phi + 
	(-\Omega_\phi) \mathbf{e}_\theta +
	\Omega_\theta \mathbf{e}_\phi \label{E:SVelocityFieldLeft}
\end{align}
where:
\begin{align}
	t_r &= \mathbf{t}\cdot \mathbf{e}_r ,&&\Omega_r = \mathbf{\Omega}\cdot \mathbf{e}_r\notag\\
	t_\theta &= \mathbf{t}\cdot \mathbf{e}_\theta ,&&\Omega_\theta = \mathbf{\Omega}\cdot \mathbf{e}_\theta\notag\\
	t_\phi &= \mathbf{t}\cdot \mathbf{e}_\phi ,&&\Omega_\phi = \mathbf{\Omega}\cdot \mathbf{e}_\phi\notag
\end{align}

The \textit{velocity field} $\mathbf{V}$ can also be expressed in spherical coordinates $(r,\theta,\phi)$ and directional unit vectors ($\mathbf{e}_r$,$\mathbf{e}_\theta$,$\mathbf{e}_\phi$):
\begin{align}
	\mathbf{V} &= \dot{r}\mathbf{e}_r + r\dot{\theta}\cos(\phi)\mathbf{e}_\theta + r\dot{\phi}\mathbf{e}_\phi \label{E:SVelocityFieldSpherical}
\end{align}
By dividing equation \eqref{E:SVelocityFieldSpherical} by $r$ we obtain:
\begin{equation}
	\frac{\mathbf{V}}{r}
	= \left(\frac{\dot{r}}{r}\right)\mathbf{e}_r + 
	\dot{\theta}\cos(\phi)\mathbf{e}_\theta + 
	\dot{\phi}\mathbf{e}_\phi\label{E:SVelocityFieldSphericalRight}
\end{equation}
We will be applying spatial partial derivatives $ \frac{\partial ()}{\partial \theta}, \frac{\partial ()}{\partial \phi}$ to equations \eqref{E:SVelocityFieldLeft} and \eqref{E:SVelocityFieldSphericalRight} using unit vector derivatives (see table \ref{Sunitvectorderivatives}). In addition we will be replacing $L$ on the resultant expressions whenever necessary given by equations \eqref{E:SLooming} and \eqref{E:SLoomingAndVelocity}.
\begin{table}[htbp]
	\caption{Partial derivatives of unit vectors $\mathbf{e}_r$,$\mathbf{e}_\theta$,$\mathbf{e}_\phi$.}
	\begin{center}
		\begin{tabular}{|c|c|c|c|}
			\hline
			& & &\\
			Unit Vector& $\frac{\partial() }{\partial \theta}$& $\frac{\partial() }{\partial \phi}$& $\frac{\partial() }{\partial r}$ \\
			& & &\\
			\hline
			$\mathbf{e}_r$& $cos(\phi)\mathbf{e}_\theta$& $\mathbf{e}_\phi$&  $0$\\
			
			\hline
			$\mathbf{e}_\theta$& $-cos(\phi)\mathbf{e}_r + sin(\phi)\mathbf{e}_\phi$& $0$&  $0$\\
			
			\hline
			$\mathbf{e}_\phi$& $-sin(\phi)\mathbf{e}_\theta$& $-\mathbf{e}_r$&  $0$\\
			
			\hline
		\end{tabular}
		\label{Sunitvectorderivatives}
	\end{center}
\end{table}
\begin{align}
	L &= -\left(\frac{\dot{r}}{r}\right)\label{E:SLooming}\\
	L &= \frac{\mathbf{t} \cdot \mathbf{e}_r}{r} \label{E:SLoomingAndVelocity}
\end{align}
By applying the partial derivatives $ \frac{\partial ()}{\partial \theta}, \frac{\partial ()}{\partial \phi}$ to equations \eqref{E:SVelocityFieldLeft} and \eqref{E:SVelocityFieldSphericalRight}, and expanding for each one of its individual components we obtain:
\begin{align}
	\frac{\partial}{\partial \theta}\left(\frac{\mathbf{V}}{r}\right) &=\frac{\partial}{\partial \theta}\left[\left(\frac{-t_r}{r}\right)\mathbf{e}_r + \left(\frac{-t_\theta}{r}\right)\mathbf{e}_\theta + 
	\left(\frac{-t_\phi}{r}\right)\mathbf{e}_\phi + 
	(-\Omega_\phi) \mathbf{e}_\theta +
	\Omega_\theta \mathbf{e}_\phi\right]\notag\\
	&=\frac{\partial}{\partial \theta}\left[\left(\frac{\dot{r}}{r}\right)\mathbf{e}_r + 
	\dot{\theta}\cos(\phi)\mathbf{e}_\theta + 
	\dot{\phi}\mathbf{e}_\phi\right] \label{E:SVPartialDerivativesTheta}
\end{align}
\begin{align}
	\frac{\partial}{\partial \phi}\left(\frac{\mathbf{V}}{r}\right) &=\frac{\partial}{\partial \phi}\left[\left(\frac{-t_r}{r}\right)\mathbf{e}_r + \left(\frac{-t_\theta}{r}\right)\mathbf{e}_\theta + 
	\left(\frac{-t_\phi}{r}\right)\mathbf{e}_\phi + 
	(-\Omega_\phi) \mathbf{e}_\theta +\Omega_\theta \mathbf{e}_\phi \right] \notag\\
	&=\frac{\partial}{\partial \phi}\left[\left(\frac{\dot{r}}{r}\right)\mathbf{e}_r + 
	\dot{\theta}\cos(\phi)\mathbf{e}_\theta + 
	\dot{\phi}\mathbf{e}_\phi\right]
	\label{E:SVPartialDerivativePhi}
\end{align}
\subsection{Partial derivatives with respect to $\theta$:}
By applying $\frac{\partial ()}{\partial \theta}$ to each one of the individual components on equation \eqref{E:SVPartialDerivativesTheta} we obtain:
\subsubsection{$\frac{\partial}{\partial \theta}\left[\left(\frac{-t_r}{r}\right)\mathbf{e}_r\right]$ component:}
\begin{align}
	\frac{\partial}{\partial \theta}\left[\left(\frac{-t_r}{r}\right)\mathbf{e}_r\right]
	&= 	\frac{\partial}{\partial \theta}\left(\frac{-t_r}{r}\right)\mathbf{e}_r + \left(\frac{-t_r}{r}\right)\frac{\partial\mathbf{e}_r}{\partial \theta}
	\notag\\	
	&= 	(-t_r)\frac{\partial}{\partial \theta}\left(\frac{1}{r}\right)\mathbf{e}_r + \left(\frac{1}{r}\right)\frac{\partial(-\mathbf{t} \cdot \mathbf{e}_r)}{\partial \theta}\mathbf{e}_r + 
	\left(\frac{-t_r}{r}\right)\cos(\phi)\mathbf{e}_\theta
	\notag\\	
	&= 	(-t_r)\left(\frac{-1}{r^2}\frac{\partial r}{\partial \theta}\right)\mathbf{e}_r + \left(\frac{1}{r}\right)\left(\frac{\partial(-\mathbf{t}) }{\partial \theta}\cdot \mathbf{e}_r\right)\mathbf{e}_r +
	\left(\frac{1}{r}\right)\left(-\mathbf{t} \cdot \frac{\partial\mathbf{e}_r}{\partial \theta}\right)\mathbf{e}_r\notag\\ &+
	\left(\frac{-t_r}{r}\right)\cos(\phi)\mathbf{e}_\theta
	\notag\\
	&= 	\frac{t_r}{r}\left(\frac{1}{r}\frac{\partial r}{\partial \theta}\right)\mathbf{e}_r + \left(\frac{1}{r}\right)\left(\cancelto{0}{\frac{\partial(-\mathbf{t}) }{\partial \theta}}\cdot \mathbf{e}_r\right)\mathbf{e}_r + cos(\phi)\left(\frac{ -\mathbf{t} \cdot \mathbf{e}\theta}{r}\right)\mathbf{e}_r \notag\\ &+
	\left(\frac{-t_r}{r}\right)\cos(\phi)\mathbf{e}_\theta
	\notag\\	
	&= \left[L\left(\frac{1}{r}\frac{\partial r}{\partial \theta}\right) + 
	\cos(\phi)\left(\frac{ -t_\theta}{r}\right)\right]\mathbf{e}_r -
	L\cos(\phi)\mathbf{e}_\theta \label{E:SfirstTheta}
\end{align}
\subsubsection{$\frac{\partial}{\partial\theta}\left[\left(\frac{-t_\theta}{r}\right)\mathbf{e}_\theta\right]$ component:}
\begin{align}
	\frac{\partial}{\partial\theta}\left[\left(\frac{-t_\theta}{r}\right)\mathbf{e}_\theta\right] 
	=&\frac{\partial}{\partial \theta}\left(\frac{-t_\theta}{r}\right)\mathbf{e}_\theta + \left(\frac{-t_\theta}{r}\right)\frac{\partial\mathbf{e}_\theta}{\partial \theta}
	\notag\\
	=& 	(-t_\theta)\frac{\partial}{\partial \theta}\left(\frac{1}{r}\right)\mathbf{e}_\theta +\left(\frac{1}{r}\right)\frac{\partial(-\mathbf{t} \cdot \mathbf{e}_\theta)}{\partial \theta}\mathbf{e}_\theta + \left(\frac{-t_\theta}{r}\right) 
	\left(-\cos(\phi)\mathbf{e}_r + \sin(\phi)\mathbf{e}_\phi\right)
	\notag\\
	=&	(-t_\theta)\left(\frac{-1}{r^2}\frac{\partial r}{\partial \theta}\right)\mathbf{e}_\theta + \left(\frac{1}{r}\right)\left(\cancelto{0}{\frac{\partial(-\mathbf{t}) }{\partial \theta}}\cdot \mathbf{e}_\theta\right)\mathbf{e}_\theta+
	\left(\frac{1}{r}\right)\left(-\mathbf{t} \cdot \frac{\partial\mathbf{e}_\theta}{\partial \theta}\right)\mathbf{e}_\theta \notag\\
	+& \cos(\phi)\left(\frac{t_\theta}{r}\right) 
	\mathbf{e}_r - \sin(\phi)\left(\frac{t_\theta}{r}\right)\mathbf{e}_\phi \notag\\
	=&\left(\frac{1}{r}\frac{\partial r}{\partial \theta}\right)\left(\frac{t_\theta}{r}\right)\mathbf{e}_\theta + \left(\frac{1}{r}\right)\left(\cos(\phi)t_r - \sin(\phi)t_\phi\right)\mathbf{e}_\theta + \cos(\phi)\left(\frac{t_\theta}{r}\right) 
	\mathbf{e}_r \notag\\ 
	-&\sin(\phi)\left(\frac{t_\theta}{r}\right)\mathbf{e}_\phi \notag\\
	=& \cos(\phi)\left(\frac{t_\theta}{r}\right)\mathbf{e}_r + \left[\left(\frac{1}{r}\frac{\partial r}{\partial \theta}\right)\left(\frac{t_\theta}{r}\right) + L\cos(\phi)- \sin(\phi)\left(\frac{t_\phi}{r}\right)\right]\mathbf{e}_\theta\notag\\
	-& \sin(\phi)\left(\frac{t_\theta}{r}\right)\mathbf{e}_\phi \label{E:SsecondTheta}
\end{align}
\subsubsection{$\frac{\partial}{\partial\theta}\left[\left(\frac{-t_\phi}{r}\right)\mathbf{e}_\phi\right]$ component:}
\begin{align}
	\frac{\partial}{\partial\theta}\left[\left(\frac{-t_\phi}{r}\right)\mathbf{e}_\phi\right] 
	=&\frac{\partial}{\partial \theta}\left(\frac{-t_\phi}{r}\right)\mathbf{e}_\phi + \left(\frac{-t_\phi}{r}\right)\frac{\partial\mathbf{e}_\phi}{\partial \theta}
	\notag\\
	=& 	(-t_\phi)\frac{\partial}{\partial \theta}\left(\frac{1}{r}\right)\mathbf{e}_\phi +\left(\frac{1}{r}\right)\frac{\partial(-\mathbf{t} \cdot \mathbf{e}_\phi)}{\partial \theta}\mathbf{e}_\phi + \left(\frac{-t_\phi}{r}\right) 
	\left(-\sin(\phi)\mathbf{e}_\theta\right)
	\notag\\
	=& (-t_\phi)\left(\frac{-1}{r^2}\frac{\partial r}{\partial \theta}\right)\mathbf{e}_\phi + \left(\frac{1}{r}\right)\left(\cancelto{0}{\frac{\partial(-\mathbf{t}) }{\partial \theta}}\cdot \mathbf{e}_\phi\right)\mathbf{e}_\phi+
	\left(\frac{1}{r}\right)\left(-\mathbf{t} \cdot \frac{\partial\mathbf{e}_\phi}{\partial \theta}\right)\mathbf{e}_\phi \notag\\
	+& \sin(\phi)\left(\frac{t_\phi}{r}\right) 
	\mathbf{e}_\theta \notag \\
	=&\left(\frac{1}{r}\frac{\partial r}{\partial \theta}\right)\left(\frac{t_\phi}{r}\right)\mathbf{e}_\phi + \left(\frac{1}{r}\right)\left(sin(\phi)t_\theta\right)\mathbf{e}_\phi + \sin(\phi)\left(\frac{t_\phi}{r}\right) 
	\mathbf{e}_\theta \notag\\
	=& \sin(\phi)\left(\frac{t_\phi}{r}\right) 
	\mathbf{e}_\theta + \left[\left(\frac{1}{r}\frac{\partial r}{\partial \theta}\right)\left(\frac{t_\phi}{r}\right) + sin(\phi)\left(\frac{t_\theta}{r}\right)\right]\mathbf{e}_\phi \label{E:SthirdTheta}
\end{align}
\subsubsection{$\frac{\partial\left(-\Omega_\phi\mathbf{e}_\theta\right) }{\partial\theta}$ component:}
\begin{align}
	\frac{\partial\left(-\Omega_\phi\mathbf{e}_\theta\right) }{\partial\theta} 
	=& \frac{\partial\left(-\Omega_\phi\right) }{\partial\theta}\mathbf{e}_\theta 	-\Omega_\phi\frac{\partial\mathbf{e}_\theta }{\partial\theta} 	\notag\\
	=& \frac{\partial\left(-\mathbf{\Omega} \cdot \mathbf{e}_\phi\right) }{\partial\theta}\mathbf{e}_\theta+ \Omega_\phi\cos(\phi)\mathbf{e}_r - \Omega_\phi\sin(\phi)\mathbf{e}_\phi\notag\\
	=& -\left(\cancelto{0}{\frac{\partial\mathbf{\Omega}}{\partial\theta}}\cdot \mathbf{e}_\phi\right)\mathbf{e}_\theta -\left(\mathbf{\Omega} \cdot  \frac{\partial\mathbf{e}_\phi }{\partial\theta}\right)\mathbf{e}_\theta+ \Omega_\phi\cos(\phi)\mathbf{e}_r - \Omega_\phi\sin(\phi)\mathbf{e}_\phi\notag\\
	=& \Omega_\theta\sin(\phi)\mathbf{e}_\theta + \Omega_\phi\cos(\phi)\mathbf{e}_r - \Omega_\phi\sin(\phi)\mathbf{e}_\phi\notag\\
	=& \Omega_\phi\cos(\phi)\mathbf{e}_r + \Omega_\theta\sin(\phi)\mathbf{e}_\theta - \Omega_\phi\sin(\phi)\mathbf{e}_\phi \label{E:SforthTheta}
\end{align}
\subsubsection{$\frac{\partial\left(\Omega_\theta\mathbf{e}_\phi\right) }{\partial\theta}$ component:}
\begin{align}
	\frac{\partial\left(\Omega_\theta\mathbf{e}_\phi\right) }{\partial\theta}
	=& \frac{\partial\Omega_\theta }{\partial\theta}\mathbf{e}_\phi 	+\Omega_\theta\frac{\partial\mathbf{e}_\phi }{\partial\theta} 	\notag\\ 
	=& \frac{\partial\left(\mathbf{\Omega} \cdot \mathbf{e}_\theta\right) }{\partial\theta}\mathbf{e}_\phi- \Omega_\theta\sin(\phi)\mathbf{e}_\theta\notag\\
	=& \left(\cancelto{0}{\frac{\partial\mathbf{\Omega}}{\partial\theta}}\cdot \mathbf{e}_\theta\right)\mathbf{e}_\phi + \left(\mathbf{\Omega} \cdot  \frac{\partial\mathbf{e}_\theta }{\partial\theta}\right)\mathbf{e}_\phi - \Omega_\theta\sin(\phi)\mathbf{e}_\theta\notag\\
	=& -\Omega_r\cos(\phi)\mathbf{e}_\phi + \Omega_\phi\sin(\phi)\mathbf{e}_\phi - \Omega_\theta\sin(\phi)\mathbf{e}_\theta\notag\\
	=& - \Omega_\theta\sin(\phi)\mathbf{e}_\theta + \left[-\Omega_r\cos(\phi) + \Omega_\phi\sin(\phi)\right]\mathbf{e}_\phi \label{E:SfifthTheta}
\end{align}
\subsubsection{$\frac{\partial}{\partial \theta}\left[\left(\frac{\dot{r}}{r}\right)\mathbf{e}_r\right]$ component:}
\begin{align}
	\frac{\partial}{\partial \theta}\left[\left(\frac{\dot{r}}{r}\right)\mathbf{e}_r\right] =& \frac{\partial}{\partial \theta}\left(\frac{\dot{r}}{r}\right)\mathbf{e}_r + \left(\frac{\dot{r}}{r}\right)\frac{\partial\mathbf{e}_r}{\partial \theta}\notag\\
	=& \dot{r}\frac{\partial}{\partial \theta}\left(\frac{1}{r}\right)\mathbf{e}_r + \left(\frac{1}{r}\right)\frac{\partial\dot{r}}{\partial \theta}\mathbf{e}_r + \left(\frac{\dot{r}}{r}\right)\cos(\phi)\mathbf{e}_\theta\notag\\
	=&\dot{r} \left(\frac{-1}{r^2}\frac{\partial r}{\partial \theta}\right)\mathbf{e}_r+ \left(\frac{1}{r}\right)\frac{\partial\dot{r}}{\partial \theta}\mathbf{e}_r + \left(\frac{\dot{r}}{r}\right)\cos(\phi)\mathbf{e}_\theta\notag\\
	=&\left[L\left(\frac{1}{r}\frac{\partial r}{\partial \theta}\right)+ \left(\frac{1}{r}\right)\frac{\partial\dot{r}}{\partial \theta}\right]\mathbf{e}_r - L\cos(\phi)\mathbf{e}_\theta \label{E:SsixTheta}
\end{align}
\subsubsection{$\frac{\partial}{\partial \theta}\left( 
	\dot{\theta}\cos(\phi)\mathbf{e}_\theta\right)$ component:}
\begin{align}
	\frac{\partial}{\partial \theta}\left( 
	\dot{\theta}\cos(\phi)\mathbf{e}_\theta\right)
	&= \cos(\phi)\frac{\partial 
		\dot{\theta}}{\partial \theta}\mathbf{e}_\theta + \dot{\theta}\cos(\phi)\frac{\partial\mathbf{e}_\theta}{\partial \theta} \notag\\
	&= \cos(\phi)\frac{\partial 
		\dot{\theta}}{\partial \theta}\mathbf{e}_\theta + \dot{\theta}\cos(\phi)(-\cos(\phi)\mathbf{e}_r + \sin(\phi)\mathbf{e}_\phi\notag\\
	&= - \dot{\theta}\cos^2(\phi)\mathbf{e}_r + \cos(\phi)\frac{\partial 
		\dot{\theta}}{\partial \theta}\mathbf{e}_\theta  + \dot{\theta}\cos(\phi)\sin(\phi)\mathbf{e}_\phi \label{E:SsevenTheta}
\end{align}
\subsubsection{$\frac{\partial}{\partial \theta}\left(
	\dot{\phi}\mathbf{e}_\phi\right)$ component:}
\begin{align}
	\frac{\partial}{\partial \theta}\left(
	\dot{\phi}\mathbf{e}_\phi\right) 
	&= 	\frac{\partial\dot{\phi}}{\partial \theta}\mathbf{e}_\phi + 	\dot{\phi}\frac{\partial\mathbf{e}_\phi}{\partial \theta}\notag\\
	&= - \dot{\phi}\sin(\phi)\mathbf{e}_\theta + \frac{\partial\dot{\phi}}{\partial \theta}\mathbf{e}_\phi \label{E:SeightTheta}
\end{align}
By adding and grouping all components \eqref{E:SfirstTheta}, \eqref{E:SsecondTheta}, \eqref{E:SthirdTheta}, \eqref{E:SforthTheta}, \eqref{E:SfifthTheta} with the corresponding components \eqref{E:SsixTheta}, \eqref{E:SsevenTheta} and \eqref{E:SeightTheta} in equation \eqref{E:SVPartialDerivativesTheta} we obtain:
\begin{flushleft}
	From $\mathbf{e}_r$ components we get:
\end{flushleft}
\begin{align}
	\cancel{L\left(\frac{1}{r}\frac{\partial r}{\partial \theta}\right) }+ 
	\cancel{\cos(\phi)\left(\frac{ -t_\theta}{r}\right) }+ \cancel{\cos(\phi)\left(\frac{t_\theta}{r}\right)} + \Omega_\phi\cos(\phi) 
	&= \cancel{L\left(\frac{1}{r}\frac{\partial r}{\partial \theta}\right)}+ \left(\frac{1}{r}\right)\frac{\partial\dot{r}}{\partial \theta} - \dot{\theta}\cos^2(\phi)\notag\\
	\Omega_\phi\cos(\phi) 
	&=\left(\frac{1}{r}\right)\frac{\partial\dot{r}}{\partial \theta} - \dot{\theta}\cos^2(\phi)\notag\\
	\Omega_\phi 
	&=\left(\frac{1}{r\cos(\phi)}\right)\frac{\partial\dot{r}}{\partial \theta} - \dot{\theta}\cos(\phi)	\label{E:SomegaPhiFromTheta}   
\end{align}
From $\mathbf{e}_\theta$ components we get:
\begin{align}
	\cancel{- L\cos(\phi)} + \left(\frac{1}{r}\frac{\partial r}{\partial \theta}\right)\left(\frac{t_\theta}{r}\right) + \cancel{L\cos(\phi)} &- \cancel{\sin(\phi)\left(\frac{t_\phi}{r}\right)}
	+ \cancel{\sin(\phi)\left(\frac{t_\phi}{r}\right)}+ \cancel{\Omega_\theta\sin(\phi)} - \cancel{\Omega_\theta\sin(\phi)} \notag\\
	&= - L\cos(\phi) + \cos(\phi)\frac{\partial 
		\dot{\theta}}{\partial \theta}  - \dot{\phi}\sin(\phi) \notag\\
	\left(\frac{1}{r}\frac{\partial r}{\partial \theta}\right)\left(\frac{t_\theta}{r}\right)  
	&= - L\cos(\phi) + \cos(\phi)\frac{\partial 
		\dot{\theta}}{\partial \theta}  - \dot{\phi}\sin(\phi) \notag\\
	L & = \frac{\partial \dot{\theta}}{\partial \theta} - \dot{\phi}\tan(\phi) - \left(\frac{1}{r}\frac{\partial r}{\partial \theta}\right)\left(\frac{t_\theta}{r\cos(\phi)}\right)\label{E:SLoomingFromTheta} 
\end{align}
And from $\mathbf{e}_\phi$ components we get:
\begin{align}
	\cancel{-\sin(\phi)\left(\frac{t_\theta}{r}\right)} &+ \left(\frac{1}{r}\frac{\partial r}{\partial \theta}\right)\left(\frac{t_\phi}{r}\right) + \cancel{sin(\phi)\left(\frac{t_\theta}{r}\right)} - \cancel{\Omega_\phi\sin(\phi)}
	-\Omega_r\cos(\phi) + \cancel{\Omega_\phi\sin(\phi)}
	\notag\\ 
	&= \dot{\theta}\cos(\phi)\sin(\phi) + \frac{\partial\dot{\phi}}{\partial \theta}\notag\\
	\Omega_r &= -\dot{\theta}\sin(\phi)- \left(\frac{1}{\cos(\phi)} \right) \frac{\partial\dot{\phi}}{\partial \theta}+ \left(\frac{1}{r}\frac{\partial r}{\partial \theta}\right)\left(\frac{t_\phi}{r\cos(\phi)}\right)\label{E:SomegaRfromTheta}
\end{align}
\subsection{Partial derivatives with respect to $\phi$:}
By applying $\frac{\partial ()}{\partial \phi}$ to each one of the individual components on equation \eqref{E:SVPartialDerivativePhi} we obtain:
\subsubsection{$\frac{\partial}{\partial \phi}\left[\left(\frac{-t_r}{r}\right)\mathbf{e}_r\right]$ component:}
\begin{align}
	\frac{\partial}{\partial \phi}\left[\left(\frac{-t_r}{r}\right)\mathbf{e}_r\right]
	&= 	\frac{\partial}{\partial \phi}\left(\frac{-t_r}{r}\right)\mathbf{e}_r + \left(\frac{-t_r}{r}\right)\frac{\partial\mathbf{e}_r}{\partial \phi}
	\notag\\	
	&= 	(-t_r)\frac{\partial}{\partial \phi}\left(\frac{1}{r}\right)\mathbf{e}_r +\left(\frac{1}{r}\right)\frac{\partial\left(\mathbf{-t} \cdot \mathbf{e}_r\right)}{\partial \phi}\mathbf{e}_r +\left(\frac{-t_r}{r}\right)\mathbf{e}_\phi
	\notag\\
	&= (-t_r)\left(\frac{-1}{r^2}\frac{\partial r}{\partial \phi}\right)\mathbf{e}_r +\left(\frac{1}{r}\right)\left(\cancelto{0}{\frac{\partial\left(\mathbf{-t}\right)}{\partial \phi}}\cdot\mathbf{e}_r\right) \mathbf{e}_r + \left(\frac{1}{r}\right)\left(\mathbf{-t} \cdot \frac{\partial\mathbf{e}_r}{\partial \phi}\right)\mathbf{e}_r + \left(\frac{-t_r}{r}\right)\mathbf{e}_\phi
	\notag\\
	&=\frac{t_r}{r}\left(\frac{1}{r}\frac{\partial r}{\partial \phi}\right)\mathbf{e}_r +  \left(\frac{-t_\phi}{r}\right)\mathbf{e}_r + \left(\frac{-t_r}{r}\right)\mathbf{e}_\phi \notag\\
	&=\left[L\left(\frac{1}{r}\frac{\partial r}{\partial \phi}\right)-\frac{t_\phi}{r}\right]\mathbf{e}_r - L\mathbf{e}_\phi\label{E:SfirstPhi}
\end{align}
\subsubsection{$	\frac{\partial}{\partial\phi}\left[\left(\frac{-t_\theta}{r}\right)\mathbf{e}_\theta\right]$ component:}
\begin{align}
	\frac{\partial}{\partial\phi}\left[\left(\frac{-t_\theta}{r}\right)\mathbf{e}_\theta\right] 
	&=\frac{\partial}{\partial \phi}\left(\frac{-t_\theta}{r}\right)\mathbf{e}_\theta + \left(\frac{-t_\theta}{r}\right)\frac{\partial\mathbf{e}_\theta}{\partial \phi}
	\notag\\
	&= (-t_\theta)\frac{\partial}{\partial \phi}\left(\frac{1}{r}\right)\mathbf{e}_\theta + \left(\frac{1}{r}\right)\frac{\partial(\mathbf{-t} \cdot \mathbf{e}_\theta)}{\partial \phi}\mathbf{e}_\theta + \left(\frac{-t_\theta}{r}\right)\cancelto{0}{\frac{\partial\mathbf{e}_\theta}{\partial \phi}}
	\notag\\
	&= (-t_\theta)\left(\frac{-1}{r^2}\frac{\partial r}{\partial \phi}\right)\mathbf{e}_\theta + \left(\frac{1}{r}\right)\left(\cancelto{0}{\frac{\partial\left(\mathbf{-t}\right)}{\partial \phi}}\cdot\mathbf{e}_\theta\right)\mathbf{e}_\theta +\left(\frac{1}{r}\right)\left(\mathbf{-t} \cdot \cancelto{0}{\frac{\partial\mathbf{e}_\theta}{\partial \phi}}\right) \mathbf{e}_\theta \notag\\
	&= \frac{t_\theta}{r}\left(\frac{1}{r}\frac{\partial r}{\partial\phi}\right)\mathbf{e}_\theta\label{E:SsecondPhi}
\end{align}
\subsubsection{$	\frac{\partial}{\partial\phi}\left[\left(\frac{-t_\phi}{r}\right)\mathbf{e}_\phi\right]$ component:}
\begin{align}
	\frac{\partial}{\partial\phi}\left[\left(\frac{-t_\phi}{r}\right)\mathbf{e}_\phi\right] 
	=&\frac{\partial}{\partial \phi}\left(\frac{-t_\phi}{r}\right)\mathbf{e}_\phi + \left(\frac{-t_\phi}{r}\right)\frac{\partial\mathbf{e}_\phi}{\partial \phi}
	\notag\\
	=&\left(-t_\phi\right)\frac{\partial}{\partial\phi}\left(\frac{1}{r}\right)\mathbf{e}_\phi + \left(\frac{1}{r}\right)\frac{\partial\left(\mathbf{-t}\cdot \mathbf{e}_\phi\right)}{\partial\phi}\mathbf{e}_\phi + \left(\frac{t_\phi}{r}\right)\mathbf{e}_r\notag\\
	=&(-t_\phi)\left(\frac{-1}{r^2}\frac{\partial r}{\partial\phi}\right)\mathbf{e}_\phi + \left(\frac{1}{r}\right)\left(\cancelto{0}{\frac{\partial(\mathbf{-t}) }{\partial\phi}}\cdot \mathbf{e}_\phi\right)\mathbf{e}_\phi + \left(\frac{1}{r}\right)\left(\mathbf{-t}\cdot\frac{\partial\mathbf{e}_\phi}{\partial\phi}\right)\mathbf{e}_\phi + \left(\frac{t_\phi}{r}\right)\mathbf{e}_r\notag\\
	=& \frac{t_\phi}{r}\left(\frac{1}{r}\frac{\partial r}{\partial\phi}\right)\mathbf{e}_\phi + \left(\frac{t_r}{r}\right)\mathbf{e}_\phi + \left(\frac{t_\phi}{r}\right)\mathbf{e}_r\notag\\
	=& \left(\frac{t_\phi}{r}\right)\mathbf{e}_r + \left[\frac{t_\phi}{r}\left(\frac{1}{r}\frac{\partial r}{\partial\phi}\right) + L\right]\mathbf{e}_\phi \label{E:SthirdPhi}
\end{align}
\subsubsection{$\frac{\partial\left(-\Omega_\phi\mathbf{e}_\theta\right) }{\partial\phi}$ component:}
\begin{align}
	\frac{\partial\left(-\Omega_\phi\mathbf{e}_\theta\right) }{\partial\phi} 
	=& \frac{\partial\left(-\Omega_\phi\right) }{\partial\phi}\mathbf{e}_\theta -\Omega_\phi\cancelto{0}{\frac{\partial\mathbf{e}_\theta }{\partial\phi}} 	\notag\\
	=& \left(\mathbf{-\Omega}\cdot\frac{\partial \mathbf{e}_\phi}{\partial \phi} \right)\mathbf{e}_\theta - \left(\cancelto{0}{\frac{\partial\mathbf{\Omega}}{\partial\phi}}\cdot\mathbf{e}_\phi\right)\mathbf{e}_\theta\notag\\
	=& \Omega_r\mathbf{e}_\theta	\label{E:SforthPhi}	
\end{align}
\subsubsection{$\frac{\partial\left(\Omega_\theta\mathbf{e}_\phi\right) }{\partial\phi}$ component:}
\begin{align}
	\frac{\partial\left(\Omega_\theta\mathbf{e}_\phi\right) }{\partial\phi}
	=& \frac{\partial\Omega_\theta }{\partial\phi}\mathbf{e}_\phi 	+\Omega_\theta\frac{\partial\mathbf{e}_\phi }{\partial\phi} 	\notag\\ 
	=& \left(\mathbf{\Omega} \cdot \cancelto{0}{\frac{\partial\mathbf{e}_\theta}{\partial\phi}}\right)\mathbf{e}_\phi + \left(\cancelto{0}{\frac{\partial\mathbf{\Omega}}{\partial\phi}} \cdot \mathbf{e}_\theta\right)\mathbf{e}_\phi - \Omega_\theta\mathbf{e}_r \notag\\
	=& -\Omega_\theta\mathbf{e}_r\label{E:SfifthPhi}
\end{align}
\subsubsection{$\frac{\partial}{\partial \phi}\left[\left(\frac{\dot{r}}{r}\right)\mathbf{e}_r\right]$ component:}
\begin{align}
	\frac{\partial}{\partial \phi}\left[\left(\frac{\dot{r}}{r}\right)\mathbf{e}_r\right] =& \frac{\partial}{\partial \phi}\left(\frac{\dot{r}}{r}\right)\mathbf{e}_r + \left(\frac{\dot{r}}{r}\right)\frac{\partial\mathbf{e}_r}{\partial \phi}\notag\\
	=& \dot{r}\frac{\partial}{\partial \phi}\left(\frac{1}{r}\right)\mathbf{e}_r + \left(\frac{1}{r}\right)\frac{\partial\dot{r}}{\partial \phi}\mathbf{e}_r + \left(\frac{\dot{r}}{r}\right)\mathbf{e}_\phi\notag\\
	=& \dot{r}\left(\frac{-1}{r^2}\frac{\partial r}{\partial\phi}\right)\mathbf{e}_r + \left(\frac{1}{r}\right)\frac{\partial\dot{r}}{\partial \phi}\mathbf{e}_r + \left(\frac{\dot{r}}{r}\right)\mathbf{e}_\phi \notag\\
	=& \left[L \left(\frac{1}{r}\frac{\partial r}{\partial\phi}\right) + \left(\frac{1}{r}\right)\frac{\partial\dot{r}}{\partial \phi} \right]\mathbf{e}_r-L\mathbf{e}_\phi\label{E:SsixPhi}
\end{align}
\subsubsection{$\frac{\partial}{\partial \phi}\left( 
	\dot{\theta}\cos(\phi)\mathbf{e}_\theta\right)$ component:}
\begin{align}
	\frac{\partial}{\partial \phi}\left( 
	\dot{\theta}\cos(\phi)\mathbf{e}_\theta\right)
	&= \frac{\partial 
		\dot{\theta}\cos(\phi)}{\partial \phi}\mathbf{e}_\theta + \dot{\theta}\cos(\phi)\cancelto{0}{\frac{\partial\mathbf{e}_\theta}{\partial \phi}} \notag\\
	&= \left[\cos(\phi)\frac{\partial\dot{\theta}}{\partial\phi} - \dot{\theta}\sin(\phi)\right]\mathbf{e}_\theta\label{E:SsevenPhi}
\end{align}
\subsubsection{$\frac{\partial}{\partial \phi}\left(
	\dot{\phi}\mathbf{e}_\phi\right)$ component:}
\begin{align}
	\frac{\partial}{\partial \phi}\left(
	\dot{\phi}\mathbf{e}_\phi\right) 
	&= 	\frac{\partial\dot{\phi}}{\partial \phi}\mathbf{e}_\phi + 	\dot{\phi}\frac{\partial\mathbf{e}_\phi}{\partial \phi}\notag\\
	&= 	\frac{\partial\dot{\phi}}{\partial \phi}\mathbf{e}_\phi - \dot{\phi} 	\mathbf{e}_r\label{E:SeightPhi}
\end{align}
By adding and grouping all components \eqref{E:SfirstPhi}, \eqref{E:SsecondPhi}, \eqref{E:SthirdPhi}, \eqref{E:SforthPhi}, \eqref{E:SfifthPhi} with the corresponding components \eqref{E:SsixPhi}, \eqref{E:SsevenPhi} and \eqref{E:SeightPhi} in equation \eqref{E:SVPartialDerivativePhi} we obtain:
\begin{flushleft}
	From $\mathbf{e}_r$ components we get:
\end{flushleft}
\begin{align}
	\cancel{L\left(\frac{1}{r}\frac{\partial r}{\partial \phi}\right)}-\cancel{\frac{t_\phi}{r}} + \cancel{\frac{t_\phi}{r}} -\Omega_\theta
	=& \cancel{L \left(\frac{1}{r}\frac{\partial r}{\partial\phi}\right)} + \left(\frac{1}{r}\right)\frac{\partial\dot{r}}{\partial \phi} - \dot{\phi} \notag\\
	\Omega_\theta
	=& -\left(\frac{1}{r}\right)\frac{\partial\dot{r}}{\partial \phi} + \dot{\phi} \label{E:SOmegaThetaFromPhi}
\end{align}
From $\mathbf{e}_\theta$ components we get:
\begin{align}
	\frac{t_\theta}{r}\left(\frac{1}{r}\frac{\partial r}{\partial\phi}\right) + \Omega_r
	=&\cos(\phi)\frac{\partial\dot{\theta}}{\partial\phi} - \dot{\theta}\sin(\phi)\notag\\
	\Omega_r
	=&- \dot{\theta}\sin(\phi)  + \cos(\phi)\frac{\partial\dot{\theta}}{\partial\phi} - \frac{t_\theta}{r}\left(\frac{1}{r}\frac{\partial r}{\partial\phi}\right)\label{E:SOmegaRFromPhi}
\end{align}
From $\mathbf{e}_\phi$ components we get:
\begin{align}
	\cancel{-L} + \frac{t_\phi}{r}\left(\frac{1}{r}\frac{\partial r}{\partial\phi}\right) + \cancel{L} &= - L + \frac{\partial\dot{\phi}}{\partial \phi}\notag\\
	L &= \frac{\partial\dot{\phi}}{\partial \phi} -\frac{t_\phi}{r}\left(\frac{1}{r}\frac{\partial r}{\partial\phi}\right)\label{E:SLoomingFromPhi} 
\end{align}
In summary, we have derived two expression for $L$, equations \eqref{E:LoomingFromTheta} and \eqref{E:SLoomingFromPhi} which we rewrite again for clarity:
\begin{align}
	L & = \frac{\partial \dot{\theta}}{\partial \theta} - \dot{\phi}\tan(\phi) - \frac{t_\theta}{r}\left(\frac{1}{\cos(\phi)}\right)\left(\frac{1}{r}\frac{\partial r}{\partial \theta}\right)\label{E:SLoomingFromTheta2}	
\end{align}	
\begin{align}
	L &= \frac{\partial\dot{\phi}}{\partial \phi} -\frac{t_\phi}{r}\left(\frac{1}{r}\frac{\partial r}{\partial\phi}\right)\label{E:SLoomingFromPhi2} 
\end{align}
Note that $\mathbf{\Omega}$ can also be calculated from the motion field and its spatial partial derivatives as shown in equations \eqref{E:SomegaRfromTheta}, \eqref{E:SOmegaRFromPhi} for $\Omega_r$, equation \eqref{E:SomegaPhiFromTheta} for $\Omega_{\theta}$, and equation \eqref{E:SOmegaThetaFromPhi} for $\Omega_{\phi}$. 
\section{Looming and Surface Normal}
Notice that the vectors $\frac{\partial \mathbf{r}}{\partial\theta}$ and $\frac{\partial \mathbf{r}}{\partial\phi}$ are both perpendicular to the surface normal vector $\mathbf{n}$ at the feature point $\mathbf{F}$. Therefore we can define the following set of constrains:
\begin{align}
	\frac{\partial r\mathbf{e}_r}{\partial\theta}\cdot\mathbf{n} &= 0\label{E:SrDotconstrainTheta}\\
	\frac{\partial r\mathbf{e}_r}{\partial\phi}\cdot\mathbf{n} &= 0\label{E:SrDotconstrainPhi}
\end{align}
By applying $\frac{\partial()}{\partial\theta}$ and $\frac{\partial()}{\partial\phi}$ to the position vector $\mathbf{r} = r\mathbf{e}_r$ in camera coordinates we get:
\subsubsection{$\frac{\partial\mathbf{r}}{\partial\theta}$ component:}
\begin{align}
	\frac{\partial r\mathbf{e}_r}{\partial\theta} &= \frac{\partial r}{\partial\theta}\mathbf{e}_r + r\frac{\partial\mathbf{e}_r}{\partial\theta}\notag\\
	&=  \frac{\partial r}{\partial\theta}\mathbf{e}_r + r\cos\phi\mathbf{e}_\theta\label{E:SrDottheta}
\end{align}
By substituting \eqref{E:SrDottheta} into the constrain \eqref{E:SrDotconstrainTheta} we obtain:
\begin{align}
	\left(\frac{\partial r}{\partial\theta}\mathbf{e}_r + r\cos\phi\mathbf{e}_\theta\right)\cdot \mathbf{n}	&= 0\notag\\
	\frac{\partial r}{\partial\theta}\left(\mathbf{e}_r\cdot\mathbf{n}\right) +  r\cos\phi\left(\mathbf{e}_\theta\cdot \mathbf{n}\right) &= 0 \notag \\
	\frac{\partial r}{\partial\theta}\left(\mathbf{e}_r\cdot\mathbf{n}\right) &=  -r\cos\phi\left(\mathbf{e}_\theta\cdot \mathbf{n}\right) \notag\\
	\frac{1}{r}\frac{\partial r}{\partial\theta} &=  -\cos\phi\left(\frac{\mathbf{e}_\theta\cdot \mathbf{n}}{\mathbf{e}_r\cdot\mathbf{n}}\right)\label{E:SrDotOverrThetaNormal}
\end{align}
\subsubsection{$\frac{\partial\mathbf{r}}{\partial\phi}$ component:}
\begin{align}
	\frac{\partial r\mathbf{e}_r}{\partial\phi} &= \frac{\partial r}{\partial\phi}\mathbf{e}_r + r\frac{\partial\mathbf{e}_r}{\partial\phi}\notag\\
	&=  \frac{\partial r}{\partial\phi}\mathbf{e}_r + r\mathbf{e}_\phi\label{E:rDotphi}
\end{align}
By substituting \eqref{E:rDotphi} into the constrain \eqref{E:SrDotconstrainPhi} we obtain:
\begin{align}
	\left(\frac{\partial r}{\partial\phi}\mathbf{e}_r + r\mathbf{e}_\phi\right)\cdot \mathbf{n}	&= 0\notag\\
	\frac{\partial r}{\partial\phi}\left(\mathbf{e}_r\cdot\mathbf{n}\right) +  r\left(\mathbf{e}_\phi\cdot \mathbf{n}\right) &= 0 \notag \\
	\frac{\partial r}{\partial\phi}\left(\mathbf{e}_r\cdot\mathbf{n}\right) &=  -r\left(\mathbf{e}_\phi\cdot \mathbf{n}\right) \notag\\
	\frac{1}{r}\frac{\partial r}{\partial\phi} &=  -\left(\frac{\mathbf{e}_\phi\cdot \mathbf{n}}{\mathbf{e}_r\cdot\mathbf{n}}\right)\label{E:SrDotOverrPhiNormal}
\end{align}
By defining surface tilt angles as $\gamma$, $\delta$:
\begin{align}
	\gamma &= \tan^{-1}\left(\frac{\mathbf{e}_\theta\cdot \mathbf{n}}{\mathbf{e}_r\cdot\mathbf{n}}\right)\label{E:StanGamma}\\
	\delta &= \tan^{-1}\left(\frac{\mathbf{e}_\phi\cdot \mathbf{n}}{\mathbf{e}_r\cdot\mathbf{n}}\right) \label{E:StanDelta}
\end{align}
We can rewrite equations \eqref{E:SrDotOverrThetaNormal}, \eqref{E:SrDotOverrPhiNormal}  as:
\begin{align}
	\frac{1}{r}\frac{\partial r}{\partial\theta} &=  -\cos\phi\tan\gamma\label{E:SrDotoverrgama}\\
	\frac{1}{r}\frac{\partial r}{\partial\phi} &=  -\tan\delta\label{E:SrDotoverrdelta}
\end{align}
In addition, we can rewrite equations \eqref{E:SLoomingFromTheta2}, \eqref{E:SLoomingFromPhi2} using \eqref{E:StanGamma}, \eqref{E:StanDelta}, \eqref{E:SrDotoverrgama} and \eqref{E:SrDotoverrdelta} as:
\begin{align}
	L &= \frac{\partial \dot{\theta} }{\partial \theta} - \dot{\phi}\tan{\phi} + \left(\frac{t_\theta}{r}\right)\tan\gamma
	\label{E:SLoominggamma}\\	
	L &= \frac{\partial \dot{\phi} }{\partial \phi} + \left(\frac{t_\phi}{r}\right)\tan\delta\label{E:SLoomingdelta}
\end{align}
Equations \eqref{E:SLoominggamma} and \eqref{E:SLoomingdelta} are essentially the same as expressions \eqref{E:SLoomingFromTheta2} and \eqref{E:SLoomingFromPhi2} but using normal notations.

\end{document}